\newcommand{\ba}{\begin{array}}
\newcommand{\ea}{\end{array}}
\newcommand{\bea}{\begin{eqnarray}}
\newcommand{\eea}{\end{eqnarray}}
\newcommand{\be}{\begin{equation}}
\newcommand{\ee}{\end{equation}}
\newcommand{\gapproxeq}{\lower .7ex\hbox{$\;\stackrel{\textstyle >}{\sim}\;$}}
\newcommand{\lapproxeq}{\lower .7ex\hbox{$\;\stackrel{\textstyle <}{\sim}\;$}}
\documentstyle[11pt]{article}
\renewcommand{\theequation}{\arabic{section}.\arabic{equation}}
\input epsf.tex
\def\pp{\pi\pi}
\def\riar{\rightarrow}
\def\mp{m_{\pi}}
\def\mk{m_{K}}
\def\fp{F_\pi}
\def\de{\delta}
\def\il{^I_l}
\def\gt{\tilde{g}}
\def\qq{\overline{q}q}
\def\1N{\displaystyle{\frac{1}{N_c}}}
\def\pe{p_\pi}
\begin{document}
\footskip 1.0cm
\thispagestyle{empty}
\setcounter{page}{0}
\baselineskip=22pt

\begin{titlepage}
%Preprint Number
\begin{flushright}
\begin{tabular}{l}
SU-4240-601\\
hep-ph/9501417\\
January 31, 1995
\end{tabular}
\end{flushright}
%TITLE
\bigskip
\begin{center}
\LARGE
\bf
Exploring $ \pp $ scattering in the $\1N$ picture
\end{center}
\vfill
%Author
\begin{center}
\Large
 {\it Francesco Sannino $^{a,b,c}$ \footnote{{\it
e-mail:}~sannino@nova.npac.syr.edu,
sannino@axpna1.na.infn.it}
{}~~{\large \it and}~~~Joseph Schechter $^{a}$ \footnote{{\it
e-mail:}~schechter@suhep.phy.syr.edu}}
\end{center}
\vfill
\begin{itemize}
\large\it
\smallskip
\item[$^a$] {Department of Physics, Syracuse University, Syracuse,
New York,
13244-1130.}

\item[$^b$] {Di\-par\-ti\-mento di Scienze Fi\-si\-che, Mo\-stra
d'Ol\-tre\-mare
Pad.\-19, 80125
Na\-po\-li, Ita\-lia.}

\item[$^c$]{ Istituto Nazionale di Fisica Nuclea\-re, Se\-zione di Na\-po\-li,
Mo\-stra
d'Ol\-tre\-mare, Pad.19 I-80125 Na\-po\-li, Italia.}
\end{itemize}
\vfill
\begin{abstract}
In the large $N_c$ approximation to $QCD$, the leading $\pp$ scattering
amplitude is expressed as the sum of an infinite number of tree diagrams.
We investigate the possibility that an adequate approximation at energies
up to somewhat more than one $GeV$ can be made by keeping diagrams which
involve the exchange of resonances in this energy range in addition to the
simplest chiral contact terms. In this approach crossing symmetry is automatic
but individual terms tend to drastically violate partial wave unitarity. We
first note that the introduction of the $\rho$ meson in a chirally invariant
manner substantially delays the onset of drastic unitarity violation which
would be present for the {\it current algebra} term alone. This suggests a
possibility of local (in energy) cancellation which we then explore in a
phenomenological way. We include exchanges of leading resonances up to
the $1.3~GeV$ region. However, unitarity requires more structure which we
model by a four derivative contact term or by a low lying scalar resonance
which is presumably subleading in the $\1N$ expansion, but may nevertheless
be important. The latter two flavor model gives a reasonable description
of the
phase shift $\delta^0_0$ up until around $860~MeV$, before the effects
associated which the $K\overline{K}$ threshold come into play.
\end{abstract}
\end{titlepage}
\newpage

\section{Introduction}
\setcounter{equation}{0}

    The study of $\pp$ scattering has been one of the classical methods
for investigating the nature of the strong interaction. Many elegant ideas have
been proposed \cite{gas}. At the present time the standard approach at low
energies
is based on chiral perturbation theory \cite{chp}. This enables one to nicely
understand
the
scattering amplitudes near threshold($<400~MeV$). However it
is very difficult to extend this treatment to higher energies since the pole
structure of the resonances in this region cannot be easily reproduced by a
truncated power series expansion in energy. Our interest in this paper
will be to investigate in a schematic sense how the description of $\pp$
scattering might be extended up to slightly past the 1 GeV region in a new
chiral picture.

   It has been clear for many years that in the region just beyond the
threshold region
 the effects of $\rho$ exchange dominate. However in the chiral
perturbation program the effects of the $\rho$ arise in the second
order of the energy expansion \cite{vmd}. This is of course due to the fact
that the usual
chiral
program is mainly devoted to improving the description of the dynamics in
the threshold region in which the $\rho$ does not explicitly appear.
 For going beyond the threshold region we would like an approach which can
treat the $\rho $ and other resonances at the first stage of an iterative
procedure.

    Such an approach is suggested by the large $ N_c $ approximation
to QCD. As reviewed in \cite{1n} for example the leading order $\1N$
approximation to $\pp$ scattering is obtained by summing all
possible tree diagrams corresponding to some effective Lagrangian
which includes an infinite number \cite{string} of bosonic resonances
of each possible spin. In addition it is allowed to include all
possible contact terms. This clearly has the right structure but
initially seems to be so general as to be practically useless. Here
we will argue that this may not be the case.

     An amplitude constructed according to the above prescription will
automatically satisfy crossing symmetry. On the other hand just calculating
the tree approximation to an effective Lagrangian will not guarantee
that unitarity is satisfied. This is the handle we will use to
try to investigate additional structure. Unitarity has of course the
consequence that the amplitude must have some suitable imaginary term which
in the usual field theory is provided by loop diagrams. However the leading
$\1N$ approximation will give a purely real amplitude away
from the singularities at the direct $s-channel$ poles. We may consider the
imaginary part of the leading $\1N$ amplitude to consist just of the
sum of delta functions at each such singularity. Clearly, the real
part has a much more interesting structure and we will mainly
confine out attention to it. Furthermore we will assume that the
singularities in the real part are {\it regularized} in a conventional way.

    Unitarity has the further
consequence that the real parts of the partial wave amplitudes must satisfy
 certain well known
bounds. The crucial question is how these bounds get satisfied since, as we
will see, individual contributions tend to violate them
 badly. At first one might expect that all of the infinite
number of resonances are really needed to obtain cancellations. However the
success
of chiral dynamics at very low energies where none of the resonances
have been taken into account suggests that this might not be the case.
At the very lowest energy the theory is described by a chiral invariant
contact interaction which, however, quite soon badly violates the unitarity
bound. It
will be observed that the $\rho$  exchange  tames this bad behavior
dramatically so that the unitarity bound is not badly broken until around an
energy beyond 2 GeV. This suggests that there is a {\it local} cancellation
between resonances which enforces the bound. The local cancellation is not
easily predicted but if true in general it would greatly simplify the
task of extending the phenomenological description of scattering processes
to higher energies. Including just the effects of the $\rho$ and the $\pi$
particles corresponds to including just the $s-wave$ quark antiquark
states in our model. A natural next step , which we shall also explore here,
would be to include the exchange of the allowed $p-wave$ quark antiquark
states. At the same time the quark model suggests that we include the first
radial excitations of the s-wave states which in fact lie near the
$p-wave$ states. Theoretically glueball and exotic states are suppressed
in $\pp$ scattering according to the large $N_c$ approximation \cite{1n}.

    In our analysis the pions will be treated as approximate Goldstone bosons
corresponding to the assumption that the theory has a spontaneously broken
chiral symmetry. Actually the need for spontaneous breakdown of this symmetry
can be argued from the $\1N$  approach itself \cite{sbs}. The effective
Lagrangian we use
shall be constructed to respect this symmetry. Furthermore for the purpose of
the initial exploration being performed here we shall consider resonance
interaction
terms with the minimum number of derivatives and shall also neglect chiral
symmetry breaking terms involving the resonances. When going beyond the initial
stage we will be forced to proceed in a more phenomenological way.

In section 2, after the presentation of the partial wave amplitudes of
interest, we show how the introduction of the $\rho-$meson in a chirally
invariant manner substantially delays the onset of the severe unitarity
violation which would be present in the simplest chiral lagrangian of pions.
The program suggested by this {\it local cancellation} is sketched.

Section 3 is concerned with the contribution to the pion scattering amplitude
from the {\it next group} of resonances - those in the range of the $p-wave$
$\qq$ bound states in the quark model. It is observed that a four derivative
contact term can be used to restore unitarity up to about $1~GeV$.

In section 4, a possibly more physical way to restore unitarity is presented
which makes use of a $\1N$ subleading contribution due to a very low mass
scalar (presumably $\qq\qq$) state.

Finally, section 5 contains a brief summary and discussion
of some directions for
future work.

%\vskip 3cm
\section{Current algebra plus $\rho$ exchange}
\setcounter{equation}{0}

In this section we will study the partial waves for $\pp$ scattering computed
in a chiral Lagrangian model which contains both the pseudoscalar and vector
mesons, (i.e., the lowest lying s-wave quark antiquark bound states).

The kinematics are discussed in Appendix A, where the partial wave amplitudes
$T\il$ are defined. They have the convenient decomposition:
\be
T\il(s)=\frac{(\eta\il (s)~e^{2i\de\il(s)}-1)}{2i}
\ee
\noindent
where $\de\il(s)$ are the phase shifts and $\eta\il(s)$ (satisfying
$0<\eta\il(s)\leq 1$) are the elasticity parameters. Extracting the real
and imaginary parts via
\bea
R\il&=&\frac{\eta\il ~sin(2\de\il)}{2},\\
I\il&=&\frac{1-\eta\il ~cos(2\de\il)}{2},
\eea
\noindent
leads to the very important bounds
\be
\big|{R\il}\big{|}\leq\frac{1}{2},~~~~~~~~~~~~~0\leq I\il\leq 1 .
\label{eq:bound}
\ee

For fixed $\eta\il$ the real and imaginary parts lie on the well known circle
in the Argand-plane $\displaystyle{{R\il}^2+(I\il-\frac{1}{2})^2
={(\frac{\eta\il}
{2})}^2}$. This formula also enables us to solve for $I\il$ as :
\be
I\il=\frac{1}{2}\left[1\pm \sqrt{{\eta\il}^2-4{R\il}^2}\right].
\label{eq:Ima}
\ee
\noindent
Let us use (\ref{eq:Ima}) for an initial orientation. Near threshold
$\eta\il=1$, $R\il$ is small and we should choose the minus sign in
(\ref{eq:Ima}) so that
\be
{I\il}(s)\approx {[R\il]}^2  .
\label{eq:approx}
\ee
\noindent
In the large $N_c$ limit the amplitude near threshold is purely real
and of the order
$\displaystyle{\frac{1}{N_c}}$. This is consistent with (\ref{eq:approx}) which
shows that $I\il(s)$ is of order $\displaystyle{\frac{1}{{N_c}^2}}$ and hence
comes in at the second order. This agrees with the chiral perturbation theory
approach \cite{chp} in which $R\il(s)$ comes from the lowest order tree diagram
while
$I\il$ arises from the next order loop diagram. On the other hand, when we
depart from the threshold region the $\displaystyle{\frac{1}{N_c}}$ approach
treats the
contribution of the $\rho$-meson at first order while the chiral perturbation
theory approach treats it at second and higher orders.

Now, pion physics at very low energies is described by the effective chiral
Lagrangian,
\be
L_{1}=-\frac{\fp^2}{8}Tr(\partial_\mu U\partial_\mu U^\dagger)
+Tr(B(U+U^\dagger))
\label{eq:algebra}
\ee
\noindent
wherein $\displaystyle{U=e^{2i\frac{\phi}{\fp}}}$ and $\phi$ is the $3\times3$
matrix of pseudoscalar fields. $\fp=132~MeV$ is the pion decay constant.
Furthermore $B=diag(B_1,B_1,B_3)$ , where $\displaystyle{B_1=\frac{\fp^2\mp^2}
{8}}$ and $\displaystyle{B_3=\frac{\fp^2(\mk^2-\frac{\mp^2}{2})}{4}}$,
describes
the minimal symmetry breaking. We shall choose $\mp=137~MeV$.

A straightforward computation using (\ref{eq:algebra}) yields the $\pp$
scattering amplitude \cite{ca} defined in (A.1):
\be
A(s,t,u)=2\frac{(s-\mp^2)}{\fp^2}.
\label{eq:Aca}
\ee
\noindent
This equation will be called the {\it current algebra result}. With
(A.2)-(A.4) we obtain $R^0_0(s)=T^0_0(s)$ as illustrated in Fig. 1.
The experimental Roy curves \cite{roy} are also shown.
Up till about $0.5~GeV$ the
agreement is quite reasonable (and can be fine tuned with second order
chiral perturbation terms) but beyond this point  $R^0_0$ keeps increasing
monotonically and badly violates the unitarity bound (\ref{eq:bound}). We will
see that the introduction of the $\rho$-meson greatly improves the situation.

There are several different but essentially equivalent ways to introduce vector
mesons into the chiral invariant Lagrangian. A simple way \cite{joe} is to
treat the
vectors as gauge particles of the chiral group and then break the local
symmetry by introducing mass-type terms. The $3\times 3$ matrix of the
vector fields, $\rho_{\mu}$ is related to auxiliary linearly transforming
gauge fields $A^L_\mu$ and $A^R_\mu$ by
\bea
A^L_\mu &=& \xi\rho_{\mu}\xi^\dagger+
\frac{i}{\gt}\xi\partial_{\mu}\xi^\dagger\\
A^R_\mu &=& \xi^\dagger\rho_{\mu}\xi+
\frac{i}{\gt}\xi^\dagger\partial_{\mu}\xi,
\eea
\noindent
where $\displaystyle{\xi\equiv U^{\frac{1}{2}}}$ and $\gt$ is a dimensionless
coupling constant. Under a chiral transformation $U\riar U_L U U_R^\dagger$
\cite{nlinear},
\be
\xi\riar U_L\xi K^\dagger\equiv K\xi U_R^\dagger
\label{eq:transf}
\ee
\noindent
(which also defines the matrix $K(\phi,U_L,U_R)$ and $\rho_\mu$ behaves as
\be
\rho_\mu\riar K\rho_{\mu}K^\dagger +\frac{i}{\gt}K\partial_{\mu}K^\dagger .
\ee
It is convenient to define
\bea
v_{\mu} &=& \frac{i}{2}\left(\xi\partial_{\mu}\xi^\dagger+
\xi^\dagger \partial_{\mu}\xi\right)\\
p_{\mu} &=& \frac{i}{2}\left(\xi\partial_{\mu}\xi^\dagger-
\xi^\dagger \partial_{\mu}\xi\right)
\label{eq:pmu}
\eea
which transform as
\bea
p_{\mu} &\riar&  Kp_{\mu}K^\dagger\\
v_{\mu} &\riar& Kv_{\mu}K^\dagger+iK\partial_{\mu}K^\dagger.
\eea

\noindent
These quantities enable us to easily construct chiral invariants and will also
 be useful later \cite{kugo}. The chiral Lagrangian including both
pseudoscalars
and vectors that one gets can be rewritten as the sum of $L_1$, in
(\ref{eq:algebra}) and the following:

\be
L_2=-\frac{1}{4}Tr(F_{\mu\nu}(\rho)F_{\mu\nu}(\rho))-
\frac{m^2_\rho}{2\gt^2}Tr\left[\left(\gt\rho_\mu-v_\mu\right)^2\right],
\label{eq:rho}
\ee
where $F_{\mu\nu}(\rho)=\partial_{\mu}\rho_{\nu}-\partial_{\nu}\rho_{\mu}-
i\gt [\rho_\mu,\rho_\nu]$. The coupling constant $\gt$ is related to the
$\rho$-meson width by
\be
\Gamma(\rho\riar2\pi)=\frac{g^2_{\rho\pp}{p_{\pi}}^3}{12\pi
m^2_\rho}~~~~~~~~~~~~~~
{}~~~g_{\rho\pp}=\frac{m^2_\rho}{\gt \fp^2}.
\ee

\noindent
We choose $m_\rho=0.769~GeV$ and $g_{\rho\pp}=8.56$. Symmetry breaking
contributions involving the $\rho$ are given elsewhere \cite{ssw} but are small
and will be neglected here. The Lagrangian piece in (\ref{eq:rho}) yields
both a pole-type contribution (from the $\rho_{\mu}v_{\mu}$ cross term)
 and
a contact term contribution (from the $v_{\mu}v_{\mu}$ term) to the
amplitude at tree level \cite{joe}:
\be
A(s,t,u)=(\ref{eq:Aca})-\frac{g^2_{\rho\pp}}{2}\left(
\frac{u-s}{m^2_\rho-t}+\frac{t-s}{m^2_\rho-u}\right)+
\frac{g^2_{\rho\pp}}{2m^2_\rho}\left[(t-s)+(u-s)\right]
\label{eq:rhoamp}
\ee
\noindent
We notice that the entire second-term in (\ref{eq:rho}) is chiral invariant
since $v_{\mu}$ and $\gt \rho_{\mu}$ transform identically. However
the $Tr(\rho_{\mu}v_{\mu})$ and $Tr(v_{\mu}v_{\mu})$ pieces are not
separately chiral invariant. This shows that the addition of the $\rho$
meson in a chiral invariant manner necessarily introduces a contact term
in addition to the minimal pole term. Adding up the terms in (\ref{eq:rhoamp})
yields finally
\be
A(s,t,u)=2\frac{(s-\mp^2)}{\fp^2}-\frac{g^2_{\rho\pp}}{2m^2_\rho}\left[
\frac{t(u-s)}{m^2_\rho-t}+\frac{u(t-s)}{m^2_\rho-u}\right]
\label{eq:rhocomp}
\ee
In this form we see that the threshold (current algebra) results
are unaffected since the second term drops out at $t=u=0$. An
alternative approach \cite{vector} to obtaining (\ref{eq:rhocomp}) involves
introducing a chiral invariant $\rho\pp$ interaction with two more
derivatives.

$A(s,t,u)$ has no singularities in the physical region. Reference to
(\ref{eq:isospin})
shows that the isospin amplitudes $T^0$ and $T^2$ also have no singularities.
However the $T^1$ amplitude has the expected singularity at $s=m^2_\rho$.
This may be cured in a conventional way, while still maintaining crossing
symmetry, by the replacements
\be
\frac{1}{m^2_\rho-t,u}\riar\frac{1}{m^2_\rho-t,u-im_\rho\Gamma_\rho}
\label{eq:propag}
\ee
in (\ref{eq:rhocomp}) \footnote{One gets a slightly different results if the
the regularization is applied to (\ref{eq:rhoamp}).}. A modification of this
sort would enter automatically
if we were to carry the computation to order $\displaystyle{\frac{1}{N^2_c}}$.
However we shall regard (\ref{eq:propag}) as a phenomenological regularization
of
the leading amplitude.

Now let us look at the actual behavior of the real parts of the partial
wave amplitudes.
$R^0_0$, as obtained from (\ref{eq:rhocomp}) with (\ref{eq:propag}),
is graphed in Fig. 2 for an extensive range of $\sqrt{s}$, together with the
{\it pions only} result from (\ref{eq:Aca}).
We immediately see that there is a remarkable improvement; the effect of
adding $\rho$ is to bend back the rising $R^0_0(s)$ so there is no longer a
drastic violation of the unitarity bound until after $\sqrt{s}=2~GeV$.
There is still a relatively small violation which we will discuss later.
Note that the modification (\ref{eq:propag}) plays no role in the improvement
since it is only the non singular $t$ and $u$ channel exchange diagrams which
contribute.

It is easy to see that the {\it delayed} drastic violation of the
unitarity bound $\displaystyle{\big|{R\il\big|}\leq\frac{1}{2}}$ is a property
of all partial waves. We have already learned from (\ref{eq:rhocomp}) that the
amplitude $A(s,t,u)$ starts out rising linearly with $s$. Now (\ref{eq:rhoamp})
and (\ref{eq:kin}) show that for large $s$ the $\rho$ exchange terms behave
as $s^0$. The leading large $s$ behavior will therefore come from the sum of
the original {\it current-algebra} term and the new {\it contact-term}:
\be
A(s,t,u)\sim\frac{2s}{\fp^2}\left(1-\frac{3k}{4}\right),~~~~~~~~~~~~~
k\equiv \frac{m^2_\rho}{\gt^2 \fp^2}.
\label{eq:rule}
\ee
But $k$ is numerically around $2$ \cite{ksrf}, so $A(s,t,u)$ eventually {\it
decreases}
linearly with $s$.
This turn-around, which is due to the contact term that enforces chiral
symmetry, delays the onset of drastic unitarity violation until well
after the $\rho$ mass. It thus seems natural to speculate that, as we go up
in energy, the leading tree contributions from the resonances we encounter (
including both crossed channel as well as $s$-channel exchange) conspire to
keep the $R\il(s)$ within the unitarity bound. We will call this possibility,
which would require that additional resonances beyond the $\rho$ come into
play when $R\il(s)$ from (\ref{eq:rhocomp}) start getting out of bound,
{\it local} cancellation.

In Fig. 3 we show the partial waves $R^1_1$ and $I^1_1$
computed using (\ref{eq:rhocomp}) and (\ref{eq:propag}). Not surprisingly,
these display the standard resonant forms.
 For completeness we present the $R^2_0$ and $R^0_2$
amplitudes in Fig. 4.
We may summarize by saying that the results of this section suggest
investigating
 the
following recipe for a reasonable approximation to the $\pp$ scattering
amplitude up to a certain scale
$E_{max}$.
\begin{itemize}
\item[1.]{Include all resonances whose masses are less than $E_{max}+\Delta$,
where $\Delta\approx$ {\it several-hundred} $MeV$.
This express the hoped for local cancellation property.}
\item[2.]{Construct all possible chiral invariants which can contribute,
presumably using the minimal number of derivatives. Compute all $\pp\riar\pp$
tree diagrams, including contact terms. {\it Regulate} the resonance
denominators in a manner similar to (\ref{eq:propag}), but restrict attention
to the real part.  Interpret the manifestly crossing symmetric result as
the leading order in $\displaystyle{\frac{1}{N_c}}$ real $\pp$ scattering
amplitude.}
\item[3.] {Obtain the imaginary parts of the partial wave amplitudes,
using (\ref{eq:Ima}). The $\eta\il(s)$ might be computed by including
channels other than $\pp$.}
\end{itemize}

We will start to explore this program by checking whether the
inclusion of the {\it next group} of resonances does enable us to satisfy the
unitary bound for $R^0_0$. For simplicity we restrict ourselves to a two-flavor
framework. It is straightforward to generalize the scheme to three flavors.

%\newpage
%\vskip 3cm
\section{The next group of resonances}
\setcounter{equation}{0}

To keep our investigation manageable we shall mainly restrict attention to the
partial wave amplitude $R^0_0(s)$. As we saw in the last section, this is the
one most likely to violate the unitarity bound. The first task is to find
the effective lagrangian which should be added to (\ref{eq:algebra}) plus
(\ref{eq:rho}). There is no {\it a priori} reason not to add chiral invariant
$\pp$ contact interactions with more than two derivatives. But the most
characteristic feature, of course, is the pionic interactions of the new
resonances in the energy range of interest, here up to somewhat more than
$1.0~GeV.$ Which resonances should be included ? In the leading
$\1N$ approximation, $\qq$ mesons are \cite{1n}
ideally mixed nonets, assuming three light quark flavors. Furthermore, the
exchange of glueballs and  exotic mesons are suppressed in interactions
with the $\qq$ mesons. The $\1N$ approximation thus directs our attention
to the $p-wave$ $\qq$ resonances as well as the radial excitations of the
$s-wave$ $\qq$ resonances.

The neutral members of the $p-wave$ $\qq$ nonets have the quantum numbers
$J^{PC}=0^{++}$, $1^{++}$, $1^{+-}$ and $2^{++}$. Of course,
the neu\-tral members
of the radially excited $s-wave$ $\qq$ nonets have
$J^{PC}=0^{-+}$ and $1^{--}$. Only members of the
$0^{++}$, $1^{--}$ and $2^{++}$ nonets can couple to two pseudoscalars
\footnote{It is
possible to write down a two point mixing interaction between $0^{-+}$ and
radially excited $0^{-+}$ particles etc.., but we shall neglect such effects
here.}. By $G$-parity conservation we finally note that it is the $I=0$
member of the $0^{++}$ and $2^{++}$ nonets and the $I=1$ member of the
$1^{--}$ nonet which can couple to two pions. Are there good experimental
candidates for these three particles ?

The cleanest case is the lighter $I=0$ member of the $2^{++}$ nonet; the
$f_2(1270)$ has, according to the August 1994 Review of Particle Properties
(RPP)
\cite{pdg}, the right quantum numbers, a mass of $1275\pm 5~MeV$, a width of
$185\pm20~MeV$, a branching ratio of $85\%$ into two pions, and a branching
ratio of only $5\%$ into $K\overline{K}$. On the other hand the
$f^{\prime}_2(1525)$ has a $1\%$ branching ratio into $\pp$ and a $71\%$
branching ratio into $K\overline{K}$. It seems reasonable to approximate
the $2^{++}$ nonet as an ideally mixed one and to regard the $f_2(1270)$ as
its non-strange member.

The $\rho(1450)$ is the lightest listed \cite{pdg} particle which is a
candidate for a radial excitation of the usual $\rho(770)$. It has a less
than $1\%$ branching ratio into $K\overline{K}$ but the $\pp$ branching ratio,
while presumably dominant, is not yet known. With this caution, we shall use
the $\rho(1450)$. The $\rho(1700)$ is a little too high for our region of
interest.

An understanding of the $I=0$, $0^{++}$ channel has been elusive despite
much work. The RPP \cite{pdg} gives
two low lying candidates: the $f_0(980)$ which has a $22\%$ branching ratio
into $K\overline{K}$ even though its central mass is below the
$K\overline{K}$ threshold and the $f_0(1300)$ which has about a $93\%$
branching ratio into $\pp$ and a $7\%$ branching ratio into
$K\overline{K}$. We shall use the $f_0(1300)$ here. It is hard to
understand why, if the $f_0(980)$ is the $\overline{s}s$ member of a
conventional $0^{++}$ nonet, it is lighter than the $f_0(1300)$.
Most likely, the $f_0(980)$ is an exotic or a $K\overline{K}$
{\it molecule} \cite{molecule}. If that is the case, its coupling to two pions
ought to be
suppressed in the $\1N$ picture. This is experimentally not necessarily true
but we will
postpone a discussion of the $f_0(980)$ as well as other possible
light $0^{++}$ resonances to the next section.

Now we will give, in turn, the $\pp$ scattering amplitudes due to the
exchange of the $f_0(1300)$, the $f_2(1270)$ and the $\rho(1450)$.
%\vskip 2cm
\subsection{The $f_0(1300)$}

Denoting a $3\times3$ matrix of scalar fields by $S$ we require that it
transform as $S\riar KSK^\dagger$, (see (\ref{eq:transf})) under, the
chiral group. A suitable chiral invariant interaction, using
(\ref{eq:pmu}), is
\be
L_{f_0}=-\gamma_0\fp^2 Tr(Sp_{\mu}p_{\mu})=-\gamma_0
Tr(S\partial_{\mu}\phi\partial_{\mu}\phi)+\cdots
\label{eq:1300}
\ee
\noindent
where the expansion of $\xi$ was used in the second step. It is interesting
to note that, in the present formalism, chiral symmetry demands that the
minimal $S\phi\phi$ interaction must have two derivatives. Specializing to
the particles of interest and taking $f_0$ to be ideally mixed leads to
\be
L_{f_0}=-\frac{\gamma_0f_0}{\sqrt{2}}
(\partial_{\mu}\vec{\pi}\cdot\partial_{\mu}\vec{\pi})+\cdots.
\label{eq:13001}
\ee
\noindent
The partial width for $f_0(1300)\riar\pp$ is then
\be
\Gamma(f_0(1300)\riar\pp)=\frac{3\gamma_0^2}{64 \pi M_{f_0}}
{\left(1-\frac{4\mp^2}{M_{f_0}}\right)}^{\frac{1}{2}}
\left(M_{f_0}^2-2\mp^2\right)^2
\label{eq:bran1300}
\ee
\noindent
The RPP\cite{pdg} lists $\Gamma_{tot}(f_0(1300))=0.15-0.40~GeV$ and
 $M_{f_0}=1.0-1.5~GeV$. For definiteness we shall choose
$\Gamma_{tot}(f_0(1300))=0.275~GeV$ and $M_{f_0}=1.3~GeV$.
These yield $|\gamma_0|=2.88~GeV^{-1}$. Using (\ref{eq:13001}) we find the
contribution of $f_0$ exchange to the $\pp$ scattering amplitude, defined
in (\ref{eq:def}), to be:
\be
A_{f_0}(s,t,u)=\frac{\gamma_0^2}{2}\frac{\left(s-2\mp^2\right)^2}
{M_{f_0}^2-s}.
\label{eq:1300ampl}
\ee
\noindent
Actually, as discussed around (\ref{eq:propag}), the singularity in the real
part of
(\ref{eq:1300ampl}) will be regulated by the replacement
\be
\frac{1}{M^2_{f_0}-s}\riar\frac{M^2_{f_0}-s}
{(M^2_{f_0}-s)^2+M_{f_0}^2\Gamma^2}.
\label{eq:1300propag}
\ee
%\vskip 2cm
\subsection{The $f_2(1270)$}

We represent the $3\times 3$ matrix of tensor fields by $T_{\mu\nu}$
(satisfying $T_{\mu\nu}=T_{\nu\mu}$, and $T_{\mu\mu}=0$) which is taken to
behave as $T_{\mu\nu}\riar KT_{\mu\nu}K^\dagger$ under chiral transformation.
A suitable chiral invariant interaction is
\be
L_T=-\gamma_2\fp^2Tr(T_{\mu\nu}p_\mu p_\nu).
\label{eq:tensor}
\ee
\noindent
Specializing to the particles of interest, this becomes
\be
L_{f_2}=-\frac{\gamma_2}{\sqrt{2}}(f_2)_{\mu\nu}
(\partial_{\mu}\vec{\pi}\cdot\partial_{\nu}\vec{\pi})+\cdots.
\label{eq:1270}
\ee
\noindent
In this case we note that the chiral invariant interaction is just the same
as the minimal one we would have written down without using chiral symmetry.
The partial width is then
\be
\Gamma(f_2(1270)\riar\pp)=\frac{\gamma_2^2}{20 \pi}\frac{p_\pi^5}{M^2_{f_2}}
\label{eq:bran1270}
\ee
\noindent
where $p_\pi$ is the pion momentum in the $f_2$ rest frame. This leads to
$|\gamma_2|=13.1~GeV^{-1}$.

To calculate the $f_2$ exchange diagram we need the spin 2 propagator
\cite{tensor}
\be
\frac{-i}{M^2_{f_2}+q^2}\left[
\frac{1}{2}\left(
\theta_{\mu_1\nu_1} \theta_{\mu_2\nu_2}+
\theta_{\mu_1\nu_2}\theta_{\mu_2\nu_1}\right)-
\frac{1}{3}\theta_{\mu_1\mu_2}\theta_{\nu_1\nu_2}\right],
\label{eq:tensorpropag}
\ee
\noindent
where
\be
\theta_{\mu\nu}=\delta_{\mu\nu}+\frac{q_\mu q_\nu}{M^2_{f_2}}.
\ee
\noindent
A straightforward computation then yields the $f_2$ contribution to the
$\pp$ scattering amplitude:
\bea
A_{f_2}(s,t,u)&=&\frac{\gamma^2_2}{2(M^2_{f_2}-s)}
\left(
-\frac{16}{3}\mp^4
+\frac{10}{3}\mp^2 s
-\frac{1}{3}s^2
+\frac{1}{2}(t^2+u^2)\right.\nonumber\\
&~&\left.-\frac{2}{3}\frac{\mp^2s^2}{M^2_{f_2}}
-\frac{s^3}{6M^2_{f_2}}
+\frac{s^4}{6M^4_{f_2}}.
\right)
\label{eq:tensorampl}
\eea
Again the singularity will be regulated as in (\ref{eq:1300propag}).

%\vskip 2cm
\subsection{The $\rho(1450)$}

We may read off the proper chiral invariant contribution\footnote{It is not
necessary to introduce the $\rho(1450)$ as a massive gauge field as we
did for the $\rho(770)$, but the answer is the same. See \cite{vector} for
further discussions.} of the
$\rho(1450)$ to the $\pp$ scattering amplitude from the second term of
(\ref{eq:rhocomp})
\be
A_{\rho^\prime}(s,t,u)=-\frac{g^2_{\rho^\prime\pp}}{2m^2_{\rho^\prime}}\left[
\frac{t(u-s)}{m^2_{\rho^\prime}-t}+\frac{u(t-s)}{m^2_{\rho^\prime}-u}\right]
\label{eq:1450ampl}
\ee
\noindent
where $g_{\rho^\prime\pp}$ is related to the $\rho^\prime\riar\pp$ partial
width by
\be
\Gamma(\rho^\prime\riar2\pi)=
\frac{g^2_{\rho^\prime\pp}\pe^3}{12\pi m^2_{\rho^\prime}}.
\ee
We shall use $|g_{\rho^\prime\pp}|=7.9$ corresponding to
$\Gamma(\rho\riar2\pi)=288~MeV$.

%\vskip 2cm
\subsection{$f_0(1300)$+$f_2(1270)$+$\rho(1450)$}
    Now we are in a position to appraise the contribution to $R^0_0$ of the
next
group of resonances. This is obtained by adding up (\ref{eq:1300ampl}),
(\ref{eq:tensorampl}) and (\ref{eq:1450ampl}) and using
(\ref{eq:isospin})$-$(\ref{eq:wave}).
The individual pieces are shown in Fig. 5.

Note that the $f_0(1300)$ piece is not the largest, as one might at first
expect. That honor goes to the $f_2$ contribution which is shown divided
into the $s$-channel pole piece and the $(t+u)$ pole piece. We observe that the
$s$-channel pole piece, associated with the $f_2$, vanishes at $\sqrt{s}=
M_{f_2}$. This happens because the numerator of the propagator in
(\ref{eq:tensorpropag}) is precisely a spin 2 projection operator at that
point. The $\rho(1450)$ contribution is solely due to the $t$ and $u$ channel
poles. It tends to cancel the $t$ and $u$ channel pole contributions of the
$f_2(1270)$ but does not quite succeed. The $t$ and $u$ channel pole
contributions of the $f_0(1300)$ turn out to be negligible. Notice the
difference in characteristic shapes of the $s$ and $(t+u)$ exchange curves.
Fig. 6 shows the sum of all these individual contributions.
There does seem to be cancellation. At the high end, $R^0_0$ starts to run
negative well
past the unitarity bound (\ref{eq:bound}) around $1.5~GeV$. But it is
reasonable to
expect resonances in the $1.5-2.0~GeV$ region to modify this. The maximum
positive value of $R^0_0$ is about $1$ at $\sqrt{s}=1.2~GeV$. This would be
acceptable if the $\pi+\rho$ contribution displayed in Fig. 2, and which must
be added to the curve of Fig. 6, were somewhat negative at this point. However
this is seen not to be the case, so some extra ingredient is required.
The $\1N$ approach still allows us the freedom of adding four, and higher
derivative contact terms. More physically, there is known to be a rather non
trivial structure below $1~GeV$ in the $I=J=0$ channel.

%\vskip 2cm
\subsection{$4-$Derivative Contact Terms.}

     First let us experiment
with four-derivative contact terms. So far we have not introduced any
arbitrary parameters but now we will be forced to do so. There are
two four-derivative chiral invariant contact interactions which are single
traces in flavor space:
\be
L_{4}=a~Tr(\partial_\mu U \partial_\nu U^\dagger
\partial_\mu U \partial_\nu U^\dagger)+
b~Tr(\partial_\mu U \partial_\mu U^\dagger
\partial_\nu U \partial_\nu U^\dagger)
\label{eq:four}
\ee
\noindent
where $a$ and $b$ are real constants. The single traces should be leading in
the $\1N$ expansion. Notice that the magnitudes of $a$ and $b$ will differ
from those in the chiral perturbation theory approach \cite{chp} since the
latter essentially also include the effects of expanding the $\rho$
exchange amplitude up to order $s^2$. The four pion terms which result from
(\ref{eq:four}) are:
\be
L_{4}=\frac{8}{\fp^4}\left[
2a\left(\partial_\mu \vec{\pi}\cdot\partial_\nu \vec{\pi}\right)^2
+(b-a)\left(\partial_\mu \vec{\pi}\cdot\partial_\mu \vec{\pi}\right)^2
\right]+
\cdots.
\label{eq:4lagr}
\ee
\noindent
This leads to the contribution to the $\pp$ amplitude:
\be
A_{4}(s,t,u)=\frac{16}{\fp^4}\left[
a\left((t-2\mp^2)^2+(u-2\mp^2)^2\right)+(b-a)(s-2\mp^2)^2\right].
\label{eq:4ampl}
\ee
\noindent
Plausibly, but somewhat arbitrarily, we will require that (\ref{eq:4ampl})
yields no correction at threshold, i.e. at $s=4\mp^2$, $t=u=0$. This gives the
condition $b=-a$ and leaves the single parameter $a$ to play with. In Fig. 7
 we show $R^0_0$, as gotten by adding the piece obtained
from (\ref{eq:4ampl}) for several values of $a$ to the contribution of
$\pi+\rho$, plus that of the {\it next group} of resonances.
For $a=+1.0\times 10^{-3}$ the
four-derivative contact term can pull the curve for $R^0_0$
down to avoid violation of the unitarity bound until around
$\sqrt{s}=1.0~GeV$. The price to be paid is that $R^0_0$ decreases very
rapidly beyond this point. We consider this to be an undesirable feature
since it would make a possible local cancellation scheme very unstable.
Another drawback of the four-derivative contact term scheme is that it lowers
$R^0_0(s)$ just above threshold, taking it further away from the Roy curves.
Let us therefore set aside the possibility of four and higher derivative
contact terms and try to find a solution to the problem of keeping $R^0_0$
within the unitarity bounds in a different and more phenomenological way.

%\vskip 3cm
\section{Low energy structure}
\setcounter{equation}{0}

Let us investigate the addition of low energy \cite{pen} {\it exotic} states
whose contributions to $\pp$ scattering should be formally suppressed
in the large $N_c$ limit. Experimentally we know that there is at least
one candidate - the $f_0(980)$ mentioned in the previous section. According to
the RPP \cite{pdg} its width is $40-400MeV$ and its branching ratio into two
pions
is $78.1 \pm 2.4\%$.
To get an idea of its effect we will choose $\Gamma_{tot}
= 40 MeV$ and use the formulas (\ref{eq:13001})-(\ref{eq:1300propag})
with the appropriate
parameters. In Fig. 8
we show the sum of the contributions of the {\it next group}
added to $\pi + \rho$ with the effect of including the $f_0(980)$.
It is seen that the $f_0(980)$ does not help unitarity - below $\sqrt{s}=
980 MeV$ it makes the situation a little worse while above it improves the
picture slightly.

    What is needed to restore unitarity over the full range of interest and
to give better agreement with the experimental data for
$\sqrt{s}\lapproxeq 900~MeV$?
\begin{itemize}
\item[{\it i.}]{Below  $450~MeV$,
$R^0_0(s)$ actually lies a little below the Roy curves.
Hence it would be nice to find a tree level mechanism which yields a small
positive addition in this region.}
\item[{\it ii.}]{In the $600-1300~MeV$ range, an
increasingly negative contribution is clearly required to keep $R^0_0$
within the unitarity bound.}
\end{itemize}

\noindent
It is possible to satisfy both of these criteria
by introducing a broad scalar resonance (like the old $\sigma$) with
a mass around $530~MeV$. Its contribution to $A(s,t,u)$ would be of the form
shown in (\ref{eq:1300ampl}) and (\ref{eq:1300propag}) which we may rewrite as:
\be
\frac{32\pi}{3H}
\frac{G}{M^3_\sigma}\frac{(s-2m_\pi^2)^2(M_\sigma^2-s)}{(s-M_\sigma^2)^2 +
M_\sigma^2{G^\prime}^2},
\label{eq:sigma}
\ee
where we have set
$\displaystyle{\frac{\gamma_0^2}{2}}\riar
\frac{32\pi}{3H}\frac{G}{M_\sigma^3}$,
 $\Gamma \riar G^\prime$ and
 $\displaystyle{H=\left(1-4\frac{\mp^2}{M^2_\sigma}\right)
 ^{\frac{1}{2}}
\left(1-2\frac{\mp^2}{M^2_\sigma}\right)^2}$ is approximately one.
If this were a typical resonance which was narrow compared to its mass and
which completely dominated the amplitude, we would set $G\approx G^\prime
\approx \Gamma$.
However for a very broad resonance it may be reasonable to regard $G^\prime$
as a phenomenological parameter which could be considered as a regulator
in the sense we have been using. Choosing $M_\sigma=0.53~GeV$,
$\displaystyle{\frac{G}{G^\prime}=0.31}$ and $G^\prime=380~MeV$, the
contribution to $R^0_0$ of
(\ref{eq:sigma}) is shown in
Fig. 9. The curve goes through zero near $0.53~GeV$ (there is a small
shift due to the crossed terms). Below this value
of $\sqrt{s}$ it adds slightly in accordance with point {\it i}
while above
$0.53~GeV$ it subtracts substantially in the manner required by point
{\it ii}.
This is the motivation behind our choice of $M_\sigma=0.53~GeV$. Adding
{\it everything} - namely the $\pi + \rho$ piece, the {\it next group} piece
together with the contributions from (\ref{eq:sigma}) and the $f_0(980)$
- results in
the curves shown in Fig. 10 for three values of $G^\prime$.
 These curves for $R^0_0(s)$ satisfy
the unitarity bound $\left|R^0_0\right|\le\frac{1}{2}$ until
$\sqrt{s}\approx 1.3~GeV$. After
$1.3~GeV$, the curves drop less precipitously than those for
the four-derivative contact term in Fig. 7.

    Fig. 10 demonstrates that the proposed {\it local cancellation} of the
various resonance exchange terms is in fact possible as a means
of maintaining
the unitarity bound for the (by construction) crossing symmetric real part of
the tree amplitude. Essentially, just the three parameters $M_\sigma,~
G$ and $G^\prime$ have been varied to obtain this. The other parameters
were all taken from experiment; when there were large experimental
uncertainties, we just selected typical values and made no attempt to
fine-tune. Procedurally, $G^\prime,~G$ and $M_\sigma$ were adjusted
to obtain a best fit to the Roy Curves below $700~MeV$; this turned out to
be what was needed for unitarity beyond $700~MeV$. It was found that
$M_\sigma$ had to lie in the $530\pm 30~MeV$ range and that
$\displaystyle{\frac{G^\prime}{G}}$ had to be in the
$0.31\pm 0.06$ range in order to achieve a fit. On the
other hand $G^\prime$ could be varied in the larger range $500\pm 315~MeV$.
It is also interesting to notice that the main effect of the sigma particle
comes from its tail in Fig. 9. Near the pole region, its effect is hidden
by the dominant $\pi+\rho$ contribution. This provides a possible
explanation of why such a state may have escaped definitive
identification.
For the purpose of comparison we show in Fig. 11, the total $R^0_0$ together
with the $\pi+\rho$ and {\it current-algebra} curves in the low energy region.

It is interesting to remark that particles with masses and widths very similar
to those above for the $\sigma$ and the $f_0(980)$ were predicted \cite{mit}
as part of a multiquark $qq\bar q\bar q$ nonet on the basis of the
$MIT$ bag model. Hence, even though
they do not give rise to formally leading $\pp$ amplitudes in the $\1N$
scheme, the picture has a good deal of plausibility from a polology
point of view. It is not hard to
imagine that some $\1N$ subleading effects might be important
at low energies where the QCD coupling constant is strongest.

Other than requiring
$\displaystyle{\big{|}R^0_0 \big{|}\leq \frac{1}{2}}$ we have
not attempted to fit the puzzling experimental results in the
the $f_0 (980)$ region. Recent interesting discussions are given in refs
\cite{pen}.
It appears that the opening of the $K\overline{K}$ channel plays an important
role and furthermore, additional resonances may be needed. In this paper we
have restricted attention to the $\pp$ channel (although the effective
Lagrangian was written down for the case of three light quarks). Clearly,
it would be interesting to study the $f_0(980)$ region
in the future, according
to the present scheme.

%\vskip 2cm
\subsection{Imaginary part and Phase Shift}

Finally, let us discuss the imaginary piece $I^0_0$. In the leading $\1N$ limit
the imaginary part vanishes away from the singularities at the poles, whereas
$R^0_0$ has support all over.
This suggests that we determine an approximation
to $I^0_0$ from the $\1N$ leading $R^0_0$ using dispersion theory, rather than
getting it directly from the tree amplitude with the regularization of the
form (\ref{eq:propag}). The latter procedure picks up pion loop
contribution to the $\rho-propagator$, for example, but misses very important
direct pion loop contributions. A dispersion approach will include both.
In the low energy region, we can proceed more simply by just using the
unitarity
formula (\ref{eq:Ima}) directly. Up until the $K\overline{K}$ threshold it
seems to be reasonable to approximate the elasticity function $\eta^0_0(s)$
by unity \cite{pen}. Strictly speaking $\eta^0_0(s)$ may depart from unity at
the
$4\pi^0$ threshold\footnote{It is amusing to note that each of the
low energy resonances, i.e. $\sigma(530)$ and $f_0(980)$, are located just
below threshold; for the $\sigma$ it is the $4\pi$ threshold while for the
$f_0(980)$ it is the $K\overline{K}$ threshold.} of $540~MeV$.
In Fig. 12 we show $I^0_0$ obtained from (\ref{eq:Ima}) on the assumption
$\eta^0_0=1$ for several values of $G^\prime$. Both signs in front of the
square root are displayed. Of course, the correct curve should start from
zero at threshold ($-$ sign in front  of the square root).
Continuity of $I^0_0(s)$ would at first appear to suggest that we follow along
the
lower curve. In order to go continuously to the upper curve it is necessary
that the argument of the square root vanish at some value of $\sqrt{s}$.
With the approximation $\eta^0_0=1$, this vanishing occurs if
$\big|R^0_0\big|$ is exactly $\displaystyle{\frac{1}{2}}$. In Fig. 12, the
discontinuity in the $\sqrt{s}=540~MeV$ region is extremely sensitive to tiny
departures of $\big|R^0_0\big|$ from $\displaystyle{\frac{1}{2}}$.
However, both experiment and the expectation that
$\displaystyle{\frac{d\delta\il}{d\sqrt{s}}\geq 0}$
\footnote{In potential theory Wigner \cite{causal} has
shown that $\displaystyle{\frac{d\delta\il}{d\sqrt{s}}\geq -\frac{a}{\beta}}$
where $a$ is the approximate interaction radius and $\beta$ is the pion
velocity in the center of mass. Strictly speaking, for $a\lapproxeq 1.7~fm$
the lower curve is also allowed.} suggest that beyond
$\sqrt{s}\approx 540~MeV$ we should actually go to the upper curve ($+$
sign in front of the square root). This can be accomplished without violating
continuity by assuming that $\eta^0_0$ is not precisely one. For the
curves shown all that is required is a decrease in $\eta^0_0$ of not more
than $0.04$. Alternatively, we could choose parameters so that $R^0_0$
reaches $0.5$ precisely. Then the fit at higher energies is slightly worse.
The corresponding three curves for the phase shifts are shown in Fig. 13.
The discontinuity should be smoothed over in accordance with our
discussion above. The agreement with experiment is quite reasonable up to
about $860~MeV$.
We did not go beyond this point for the purpose of obtaining the phase shift
because we are neglecting the $K\overline{K}$ channel which becomes relevant
in the computation of the imaginary part.

%\newpage
%\vskip 3cm
\section{Summary and discussion}
\setcounter{equation}{0}

In the leading large $N_c$ approximation to $QCD$, $\pp$ scattering corresponds
to the sum of an infinite number of tree diagrams which can be of the contact
type or can involve resonance exchange. This can only be a practically useful
approximation if it is possible to retain just a reasonably small number of
terms. The most natural way to do so is, of course, to consider contact terms
with as few derivatives as possible and exchange terms with resonances
having masses less than the extent of the energy region we wish to describe.
In this paper we have carried out an initial exploration of this program in
a step by step way. The first step is to include only the well known chiral
contact term which reasonably describes the scattering lengths. However this
amplitude badly violates partial wave unitarity bounds (seen most readily in
the $I=L=0$ channel, see Fig. 1) at energies beyond $500~MeV$. We observed that
the introduction of the $\rho$ meson dramatically improved the situation,
delaying drastic violation of the unitarity bound till around $2~GeV$
(see Fig. 2).
We noted that this effect could be nicely understood as the result of an extra
contact term which must be present when the $\rho$ is introduced in a chirally
invariant manner. Furthermore, this feature holds in the strict large $N_c$
limit, i.e., without including the phenomenological regularization
(\ref{eq:propag}). The observed cancellation encouraged us to investigate the
possibility of a more general {\it local cancellation}, due to inclusion
of all (large $N_c$ leading) resonances in the energy range of interest and,
perhaps, higher derivative contact terms.
The program is sketched at the end of section 2.

Taking the large $N_c$ approach as well as the standard $\qq$ spectrum
literally, we argued that the {\it next group} of resonances whose exchange
contributes to the leading amplitude should comprise the $f_0(1300)$, the
$f_2(1270)$ and the $\rho(1450)$. We observed (section 3) that there was a
tendency for these to cancel among themselves; for example the crossed-channel
exchanges of the $\rho(1450)$ tended to cancel against those of the
$f_2(1270)$. In our analysis, the complications due to enforcing chiral
symmetry
and using the full spin 2 propagator were taken into account. However, the
cancellation with both the $\pi+\rho$ and {\it next group} was not sufficient
to satisfy the unitarity bound
$\displaystyle{\big|R^0_0\big|\le\frac{1}{2}}$ in the energy range till
$1.3~GeV$. An allowed leading $N_c$ way out - by adding four derivative contact
terms - was thus investigated. This enabled us to restore unitarity till about
$1.0~GeV$ (see Fig. 7). The drawback was that $R^0_0(s)$ dropped off rather
precipitously afterwards, which would make a local cancellation scheme very
unstable.

As a more physically motivated alternative we investigated, in section 4,
the possibility of including scalar resonances having masses less than
$1~GeV$. These are presumably not of the simple $\qq$ type and hence their
exchange should be of sub-leading order in the large $N_c$ limit. An
interesting
interpretation gives these particles a $\qq\qq$ quark structure \cite{mit}.
Then a somewhat narrow state like the $f_0(980)$ is expected together with
a very low mass and very broad state like the old $\sigma$ meson.
(Both should belong to a $3-$flavor nonet). It was found that the $f_0(980)$
particle did not help much in restoring unitarity. In the experimentally
puzzling region close to $980~MeV$ it is, however, expected to play an
extremely important role. On the othe hand, the further introduction of the
other
scalar, which we denoted as the $\sigma(530)$, treated with a phenomenological
regularization parameter (see (\ref{eq:sigma})) enabled us to satisfy the
unitarity bound all the way up to $1.3~GeV$ (see Fig. 10). Thus, if low
energy scalars are included, the proposed {\it local cancellation} may
be a viable possibility. The imaginary part of the partial wave amplitude,
$I^0_0(s)$ was also computed from the unitarity relation (\ref{eq:Ima}) and
found to lead to a phase shift $\delta^0_0$ in reasonable agreement with
experiment until about $860~MeV$. Beyond this point, the effect of the
opening of the $K\overline{K}$ channel must be specifically included.

There are many directions for further work.
\begin{itemize}
\item[{\it i.}]{The most straightforward is the investigation of different
channels. For example, considering $\pp\riar K\overline{K}$ and
$K\overline{K}\riar K\overline{K}$ should enable us to study the
interesting $K\overline{K}$ threshold region in a more detailed way. Looking
at channels which don't communicate with $\pp$ would enable one to focus
on particular resonance exchanges.}
\item[{\it ii.}] {The greatly increasing density of levels as one goes
up in energy clearly indicates that there is a limit to how far one can go
with the kind of {\it microscopic} approach presented here. It is expected
that at energies not too much higher than the $1.3~GeV$ region this
analysis should merge with some kind of string-like picture \cite{string}.
In that region the question of the validity of the $\1N$ expansion and a
possible {\it local cancellation} can presumably be approached in a more
analytical manner and interesting models can be studied \cite{lnc}. Here we
have tried
to follow a phenomenologically oriented path, assuming only chiral dynamics
in addition to the $\1N$ framework.}
\item[{\it iii.}] {One can also imagine a kind of {\it Wilsonian effective
action} \cite{pheno} with which the present approach can be further discussed.
This should allow the systematic calculation of loops but would be extremely
complicated in practice.}
\end{itemize}

%\vskip 2cm

\begin{center}
{\bf Acknowledgments}
\end{center}
We would like to thank Masayasu Harada for helpful discussions. One
of us (F.S.) would like to thank Prof. R. Musto for pointing out the
possible relevance of ref. \cite{pheno}. This work
was supported in part by the U.S. DOE Contract No. DE-FG-02-85ER40231.
\newpage
\section*{Appendix A}
\setcounter{equation}{0}
\renewcommand{\theequation}{A.\arabic{equation}}

    Here we list the kinematic conventions for $\pp$ scattering. The invariant
amplitude for $ \pi_i + \pi_j \riar  \pi_k + \pi_l $ is decomposed as:
\be
 \delta_{ij}\delta_{kl} A(s,t,u) + \delta_{ik}\delta_{jl} A(t,s,u)
+ \delta_{il}\delta_{jk} A(u,t,s)  ,
\label{eq:def}
\ee
where s,t and u are the usual Mandelstam variables obeying $ s+t+u=4\mp^2 $.
Physical values lie in the region  $s{\geq}4\mp^2,~t{\leq}
0, u {\leq}0$. (Note that the phase of (\ref{eq:def}) corresponds to simply
taking
the matrix element of the Lagrangian density of a four point contact
interaction).

   Projecting out amplitudes of definite isospin yields:
\bea
T^0(s,t,u) &=& 3A(s,t,u)+A(t,s,u)+A(u,t,s),\nonumber\\
T^1(s,t,u) &=& A(t,s,u)-A(u,t,s),\nonumber\\
T^2(s,t,u) &=& A(t,s,u)+A(u,t,s).
\label{eq:isospin}
\eea

    In the center of mass frame:
\bea
   s=4(\pe^2+{\mp}^2),\nonumber\\
   t=-2\pe^2(1-cos\theta),\nonumber\\
   u=-2\pe^2(1+cos\theta),
\label{eq:kin}
\eea
where $\pe$ is the spatial momentum and $\theta$ is the scattering angle. We
then
define
the partial wave isospin amplitudes  according to the following formula:
\be
T^{I}_{l}(s)\equiv \frac{1}{64\pi} \sqrt{\left(1-4\frac{\mp^2}{s}\right)}
\int^{1}_{-1}dcos
\theta
P_l(cos\theta) T^I(s,t,u).
\label{eq:wave}
\ee

\newpage
\begin{center}
\epsfxsize=400pt
\ \epsfbox{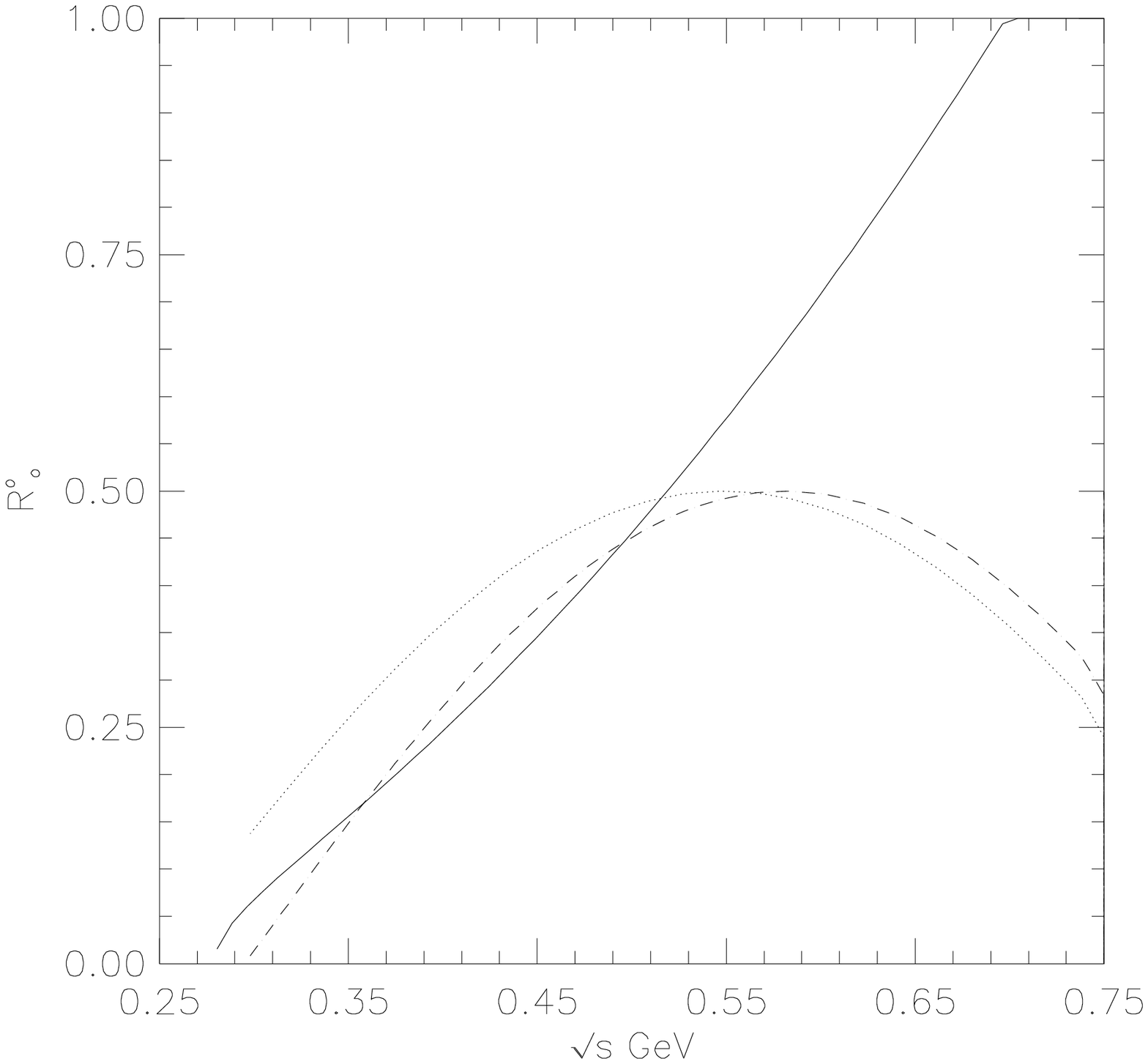}
\begin{itemize}
\item[Fig. 1]{The solid line is the current algebra result for $R^0_0$.
The dotted and
dot-dashed lines are the Roy curves for $R^0_0$.}
\end{itemize}
\end{center}
\noindent

\newpage
\begin{center}
\epsfxsize=400pt
\ \epsfbox{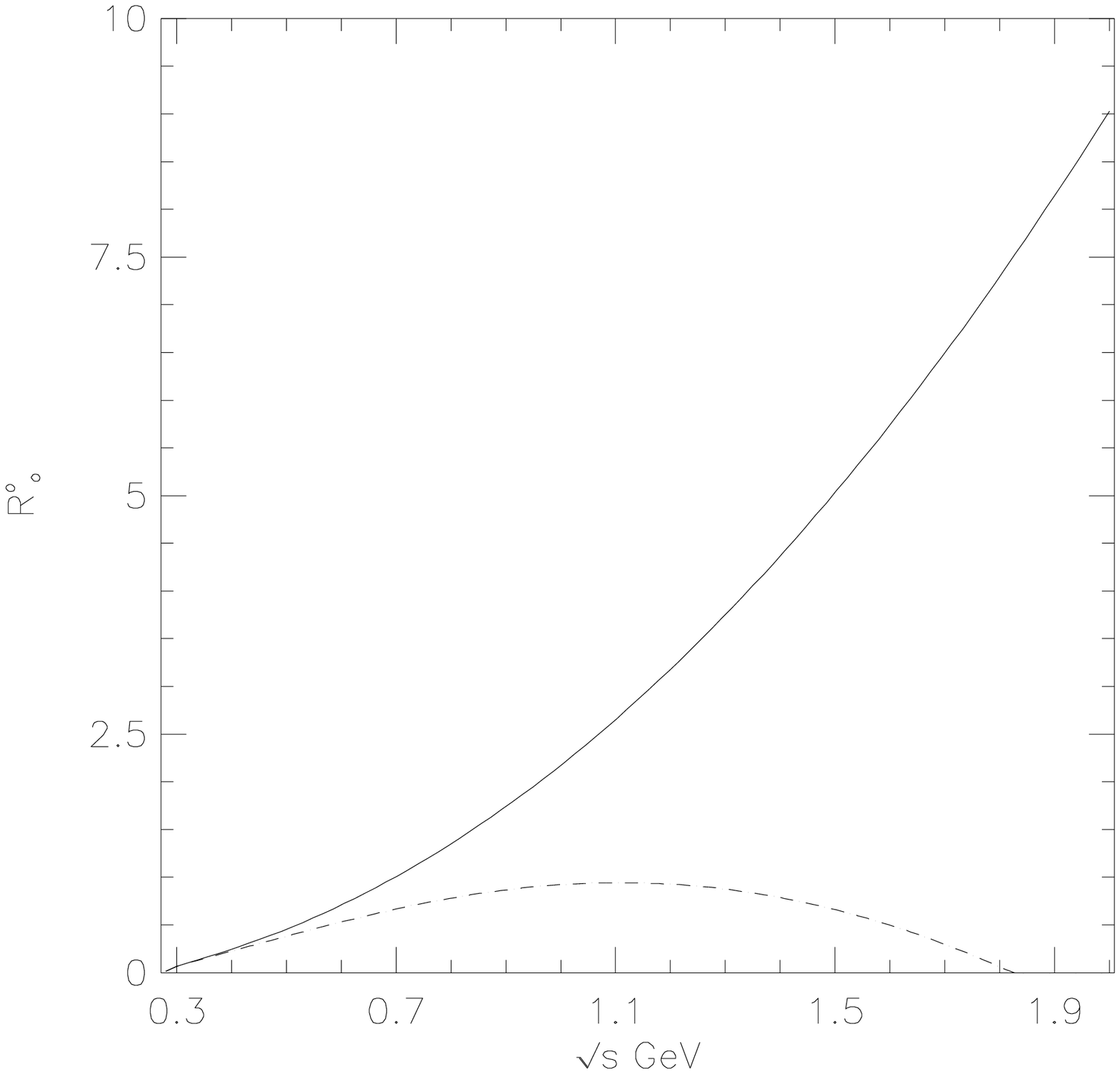}
\begin{itemize}
\item[Fig. 2]{The solid line is the current algebra result for $R^0_0$.
The dot-dashed line is the $\rho+\pi$ result for $R^0_0$.}
\end{itemize}
\end{center}
\noindent
\newpage

\begin{center}
\epsfxsize=400pt
\ \epsfbox{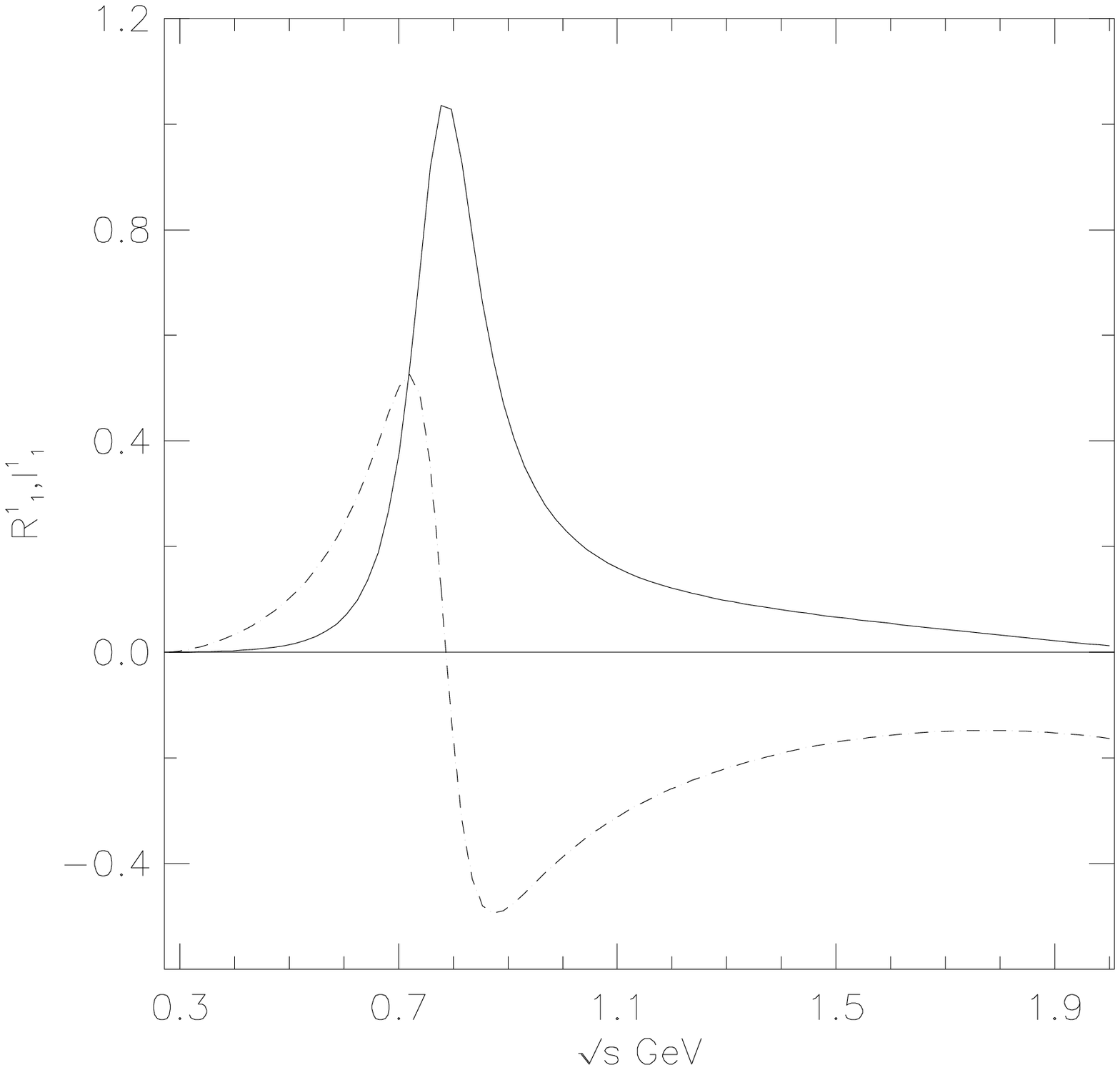}
\begin{itemize}
\item[Fig. 3]{The solid line is the imaginary part $I^1_1$.
The dot-dashed line is $R^1_1$.}
\end{itemize}
\end{center}
\noindent
\newpage

\begin{center}
\epsfxsize=400pt
\ \epsfbox{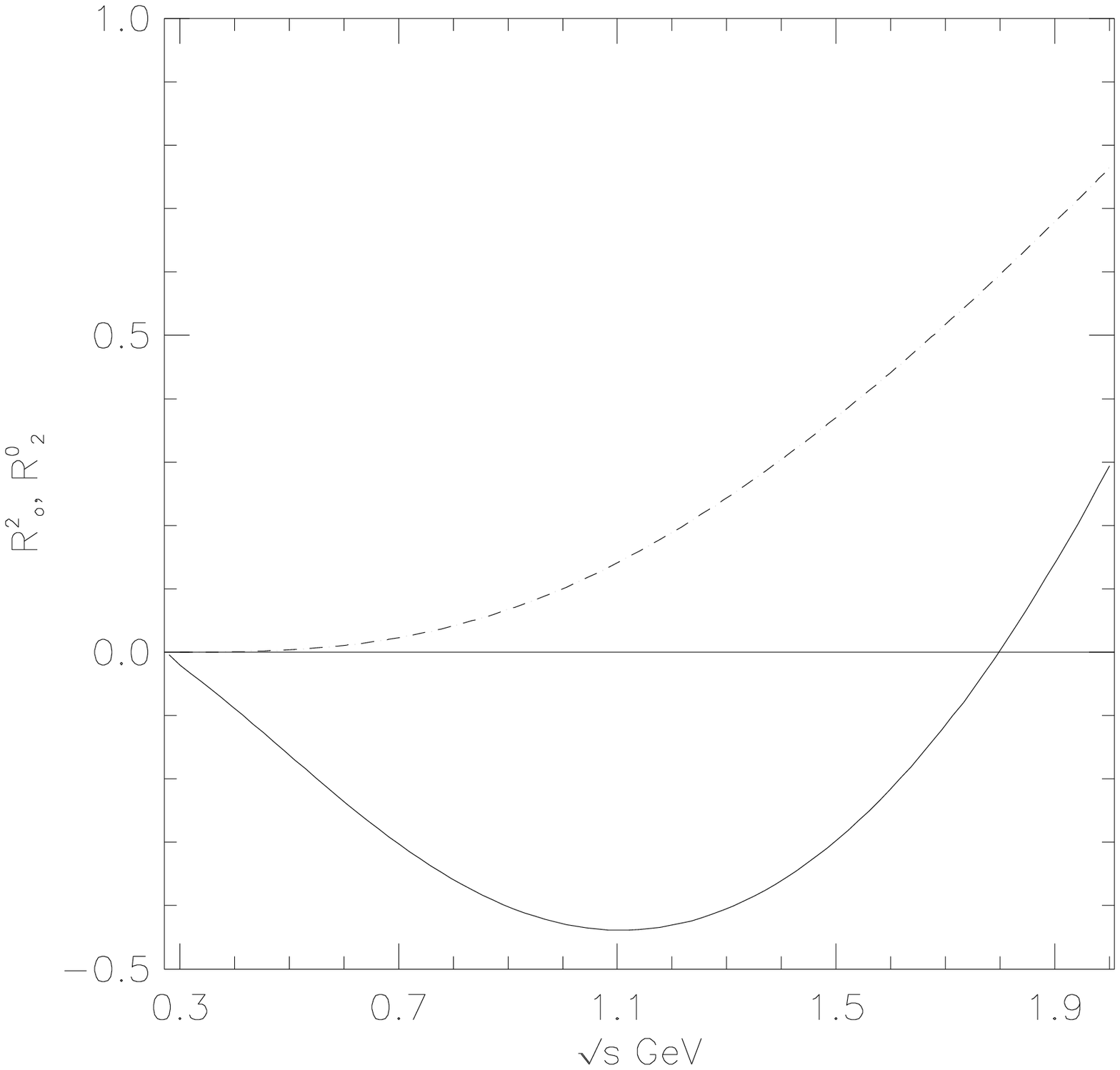}
\begin{itemize}
\item[Fig. 4]{The solid line is the $\pi+\rho$ contribution
 to the $I=2,~L=0$ real part.
The dot-dashed line is the $\pi+\rho$ contribution
 to the $I=0, L=2$ real part.}
\end{itemize}
\end{center}
\newpage

\begin{center}
\epsfxsize=400pt
\ \epsfbox{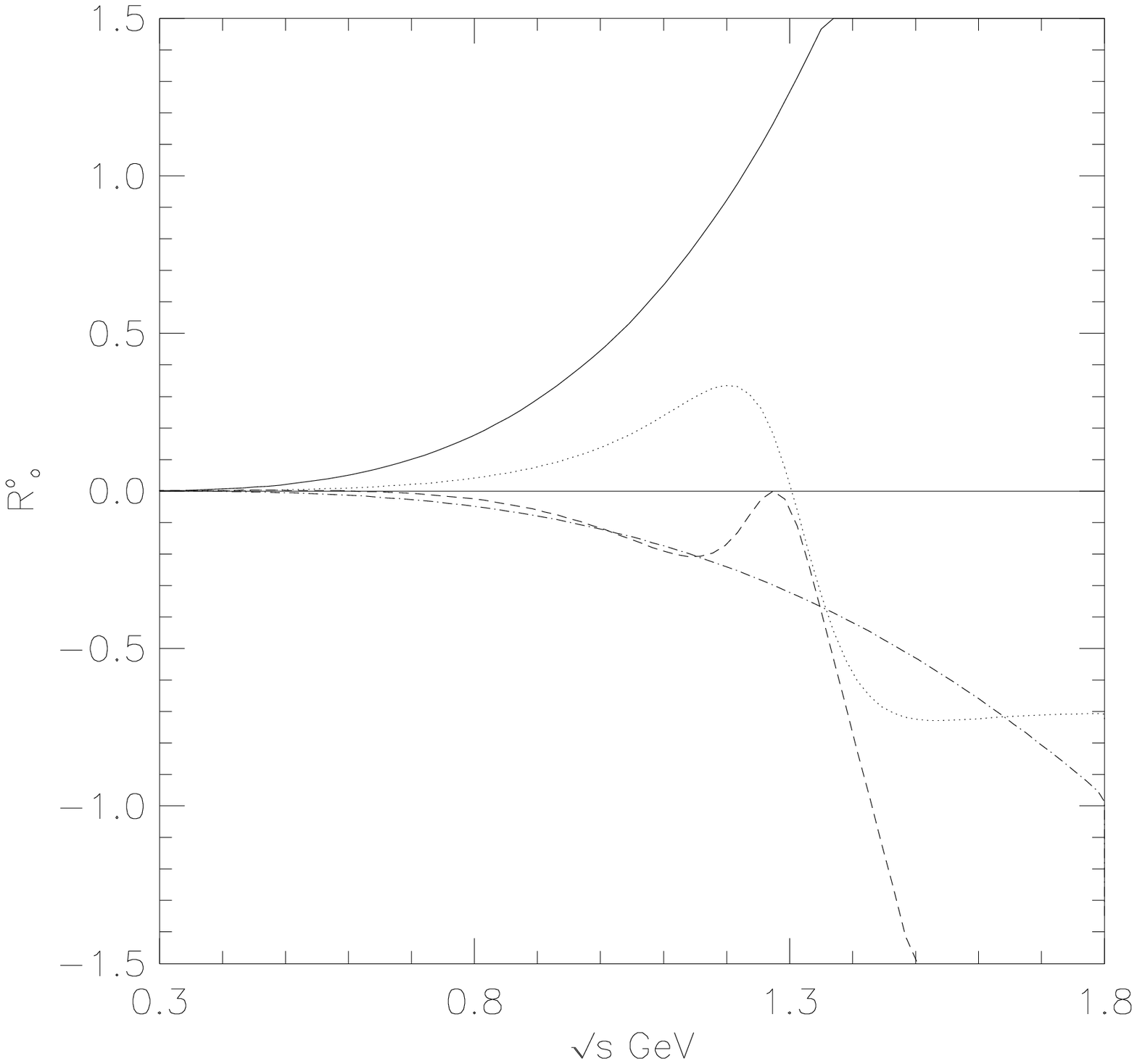}
\begin{itemize}
\item[Fig. 5]{Contributions to $R^0_0$. Solid line: $f_2(t+u)$. Dashed line:
$f_2(s)$. Dotted line:
$f_0(1300)$. Dot-dashed line: $\rho(1450)$.}
\end{itemize}
\end{center}
\noindent
\newpage

\begin{center}
\epsfxsize=400pt
\ \epsfbox{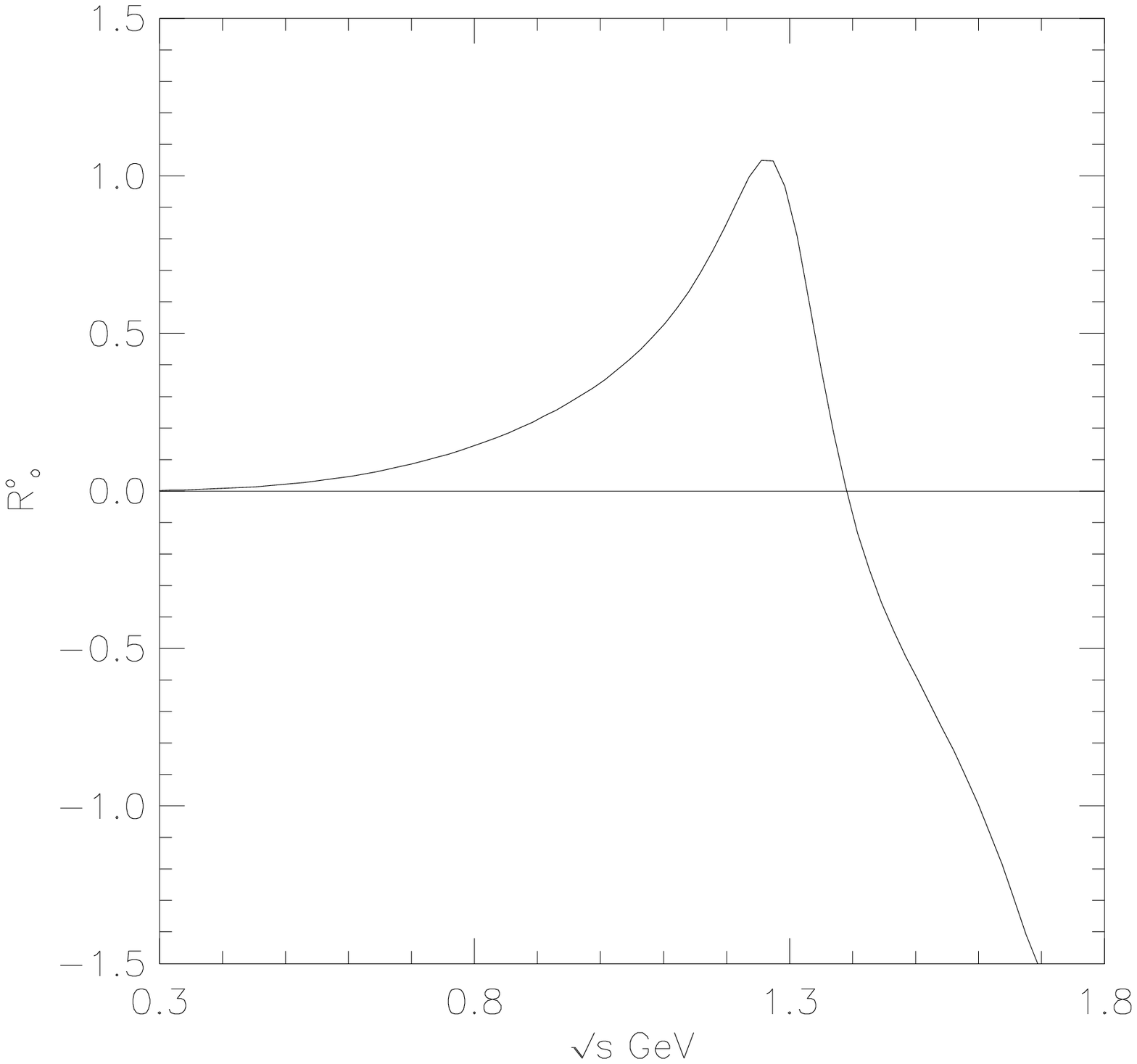}
\begin{itemize}
\item[Fig. 6]{Sum of all contributions in Fig. 5.}
\end{itemize}
\end{center}
\noindent
\newpage

\begin{center}
\epsfxsize=400pt
\ \epsfbox{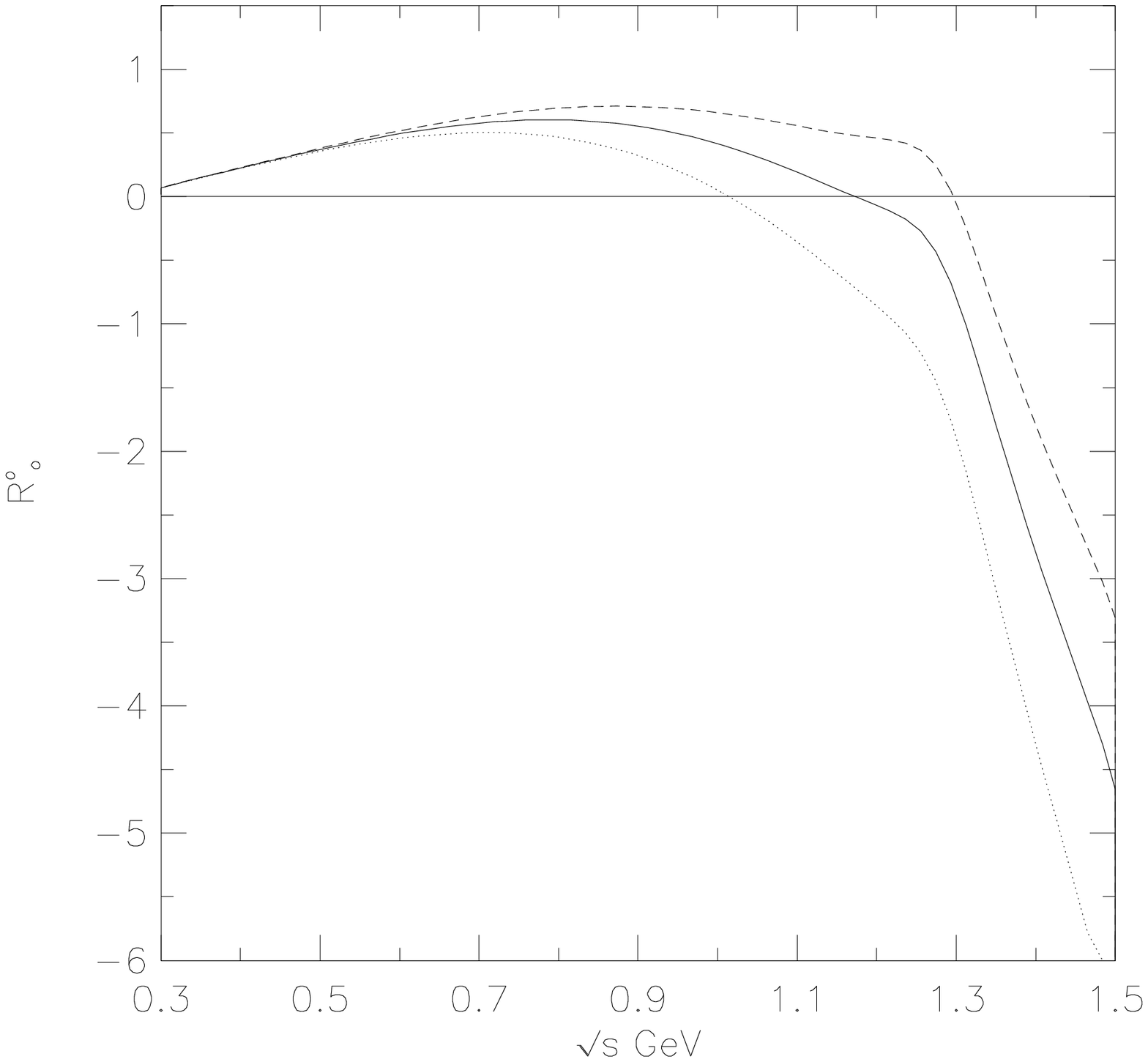}
\begin{itemize}
\item[Fig. 7]{Effect of four derivative contact term.
Solid line: $a=+0.7$. Dashed line: $a=+1.0$. Dotted line: $a=+0.5$, in
units of $10^{-3}$ }
\end{itemize}
\end{center}
\noindent
\newpage

\begin{center}
\epsfxsize=400pt
\ \epsfbox{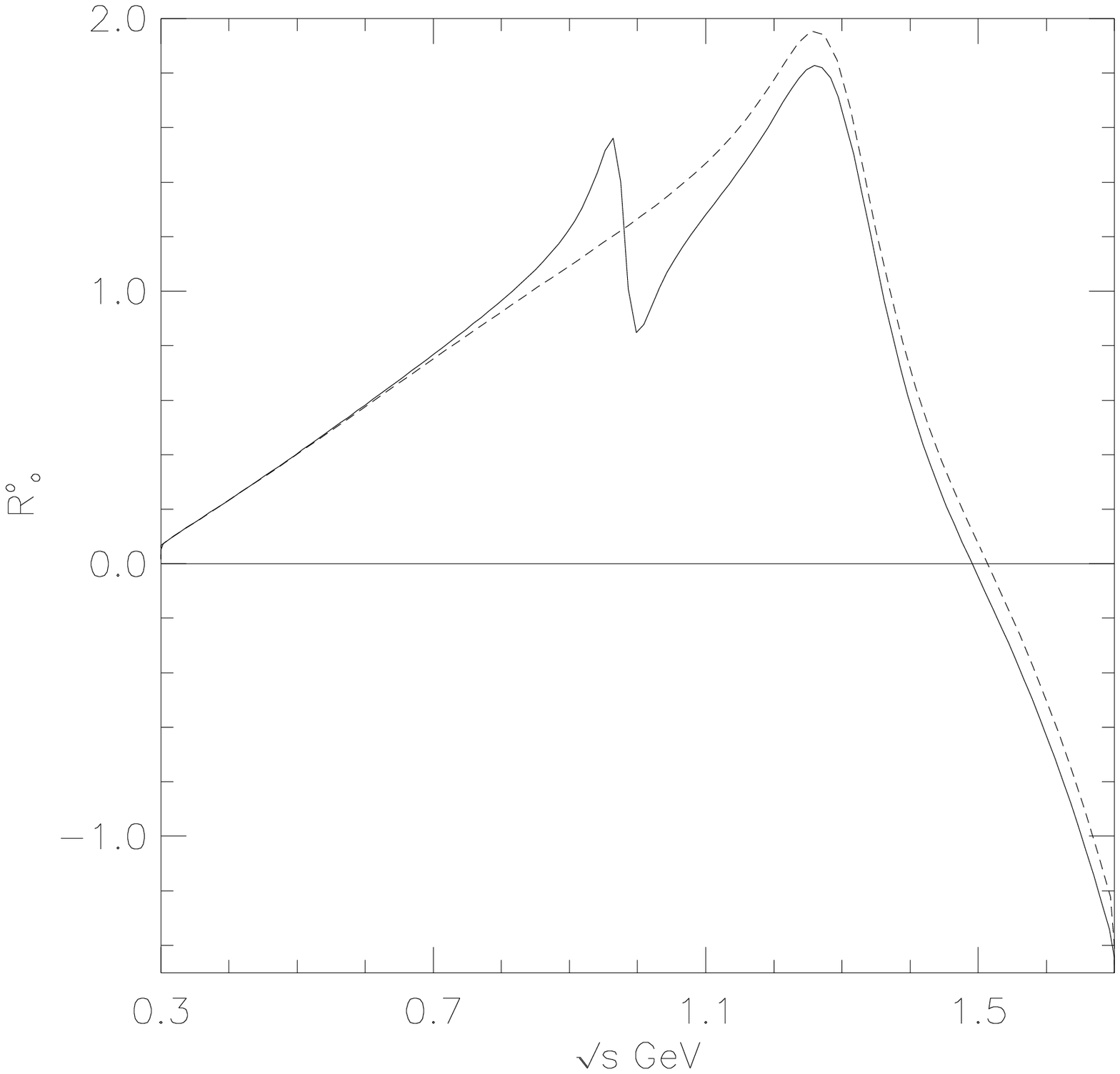}
\begin{itemize}
\item[Fig. 8]{Solid line: $\pi+\rho(770)+f_0(980)+f_2(1275)+f_0(1300)
+\rho(1450)$. Dashed line: without $f_0(980)$.}
\end{itemize}
\end{center}
\noindent
\newpage

\begin{center}
\epsfxsize=400pt
\ \epsfbox{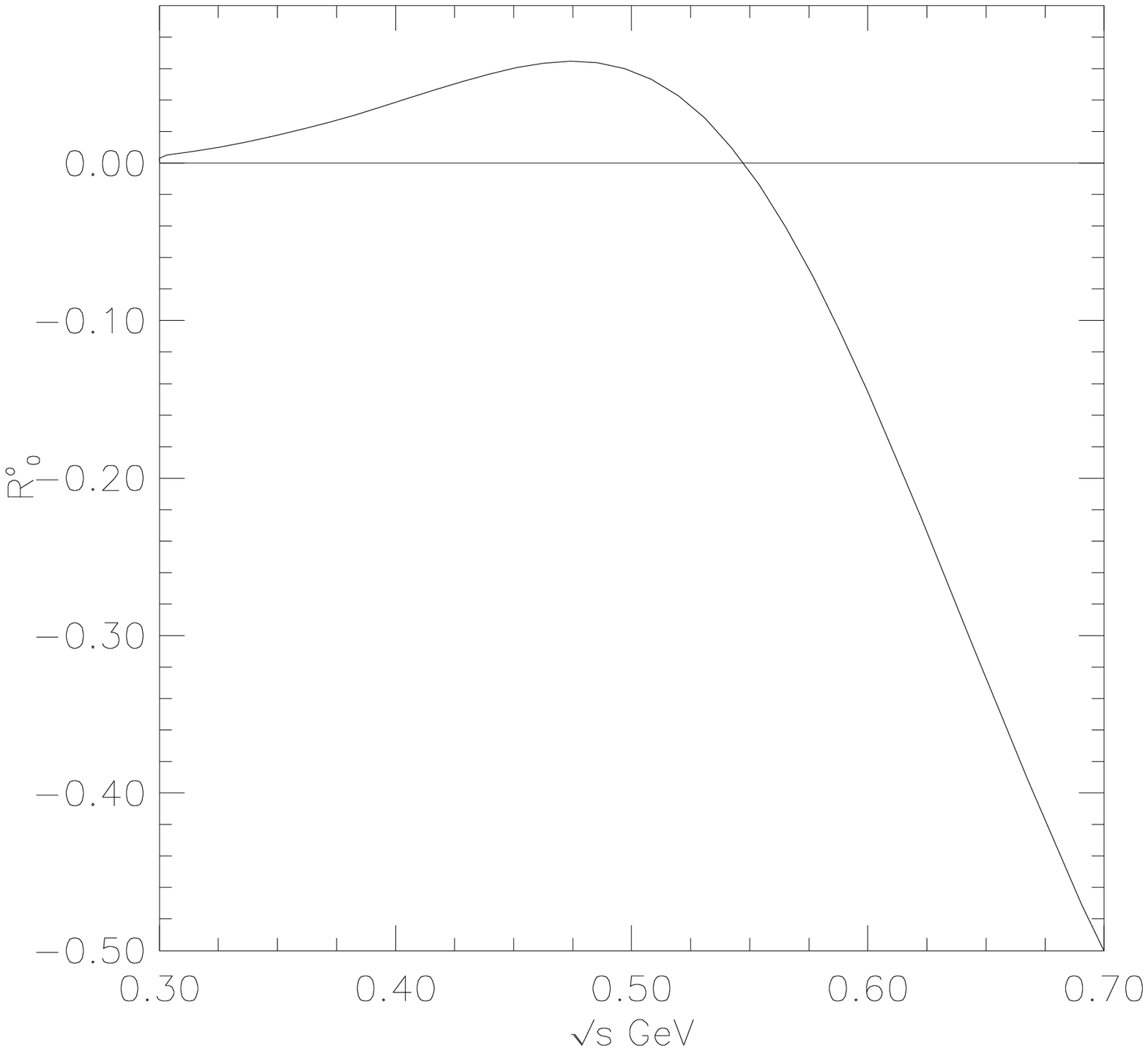}
\begin{itemize}
\item[Fig. 9]{Contribution of $\sigma(530)$ to $R^0_0$.}
\end{itemize}
\end{center}
\noindent
\newpage

\begin{center}
\epsfxsize=400pt
\ \epsfbox{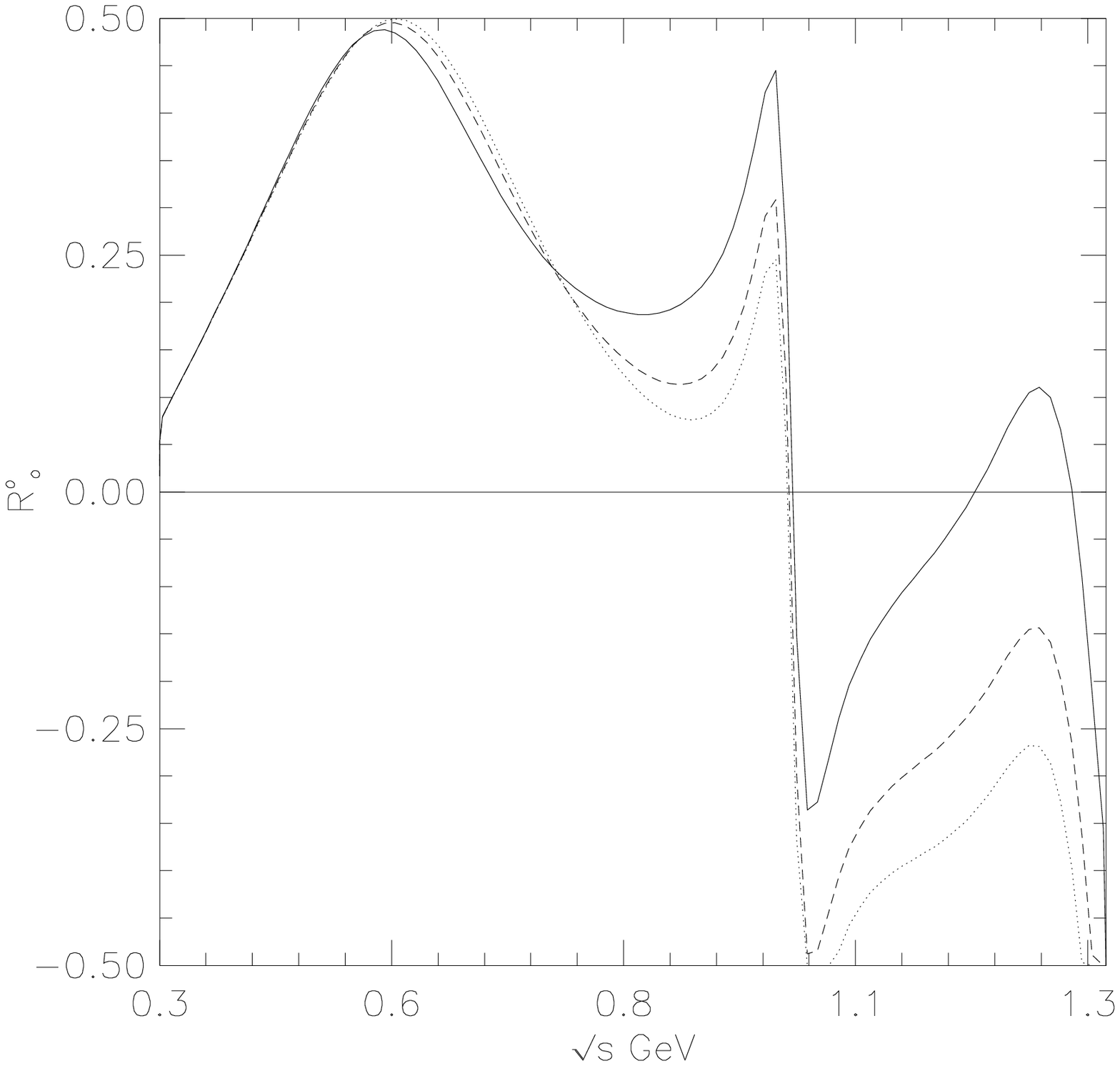}
\begin{itemize}
\item[Fig. 10]
{Pattern for $R^0_0$. Solid line: $G^\prime=380~MeV$. Dashed line:
$G^\prime=440~MeV$.
Dotted line: $G^\prime=470~MeV$.}
\end{itemize}
\end{center}
\noindent
\newpage

\begin{center}
\epsfxsize=400pt
\ \epsfbox{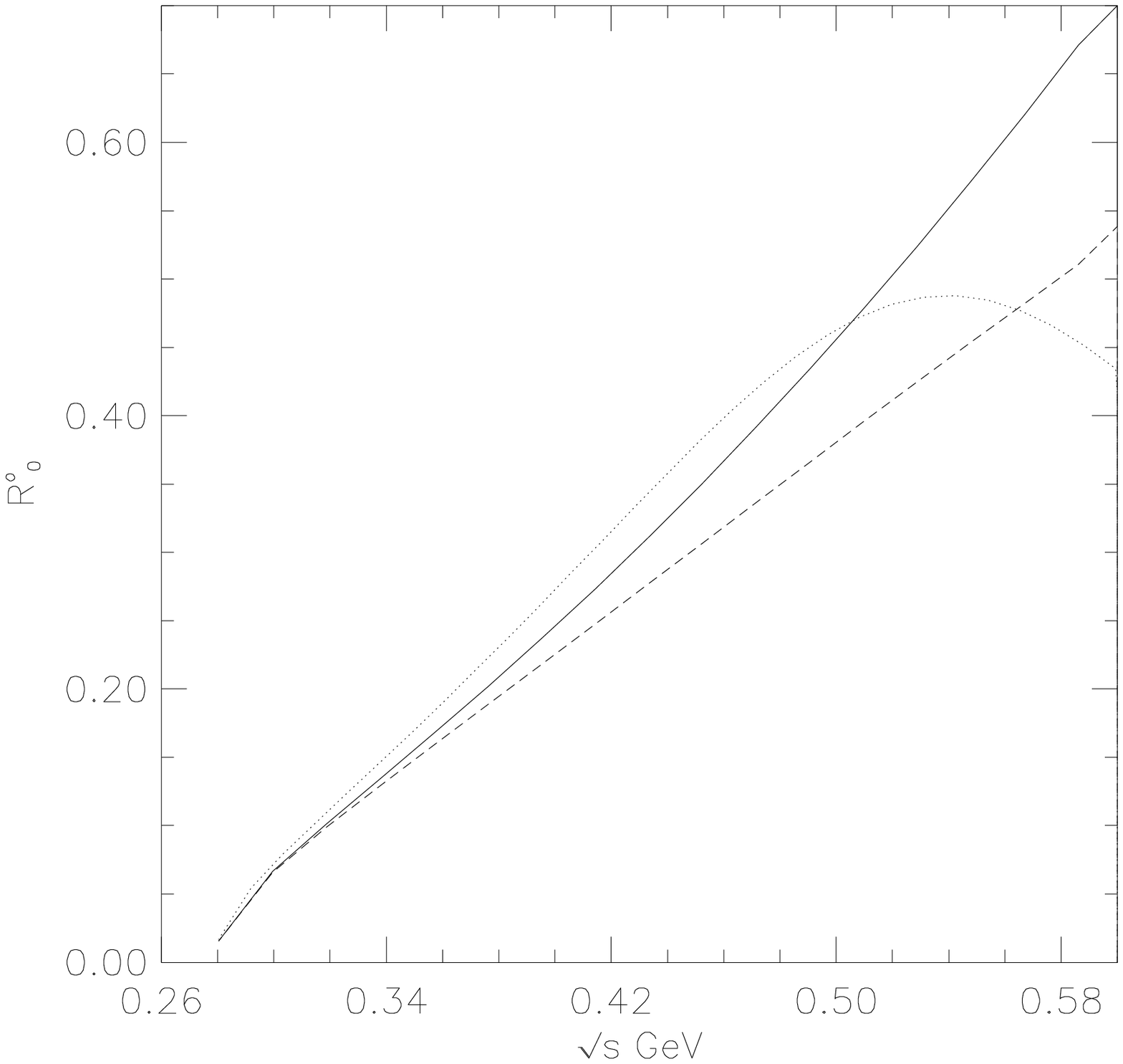}
\begin{itemize}
\item[Fig. 11]{The low energy structure for
{\it current-algebra}, solid-line; $\pi+\rho$, dashed-line;
{\it everything}, dotted-line.}
\end{itemize}
\end{center}
\noindent
\newpage

\begin{center}
\epsfxsize=400pt
\ \epsfbox{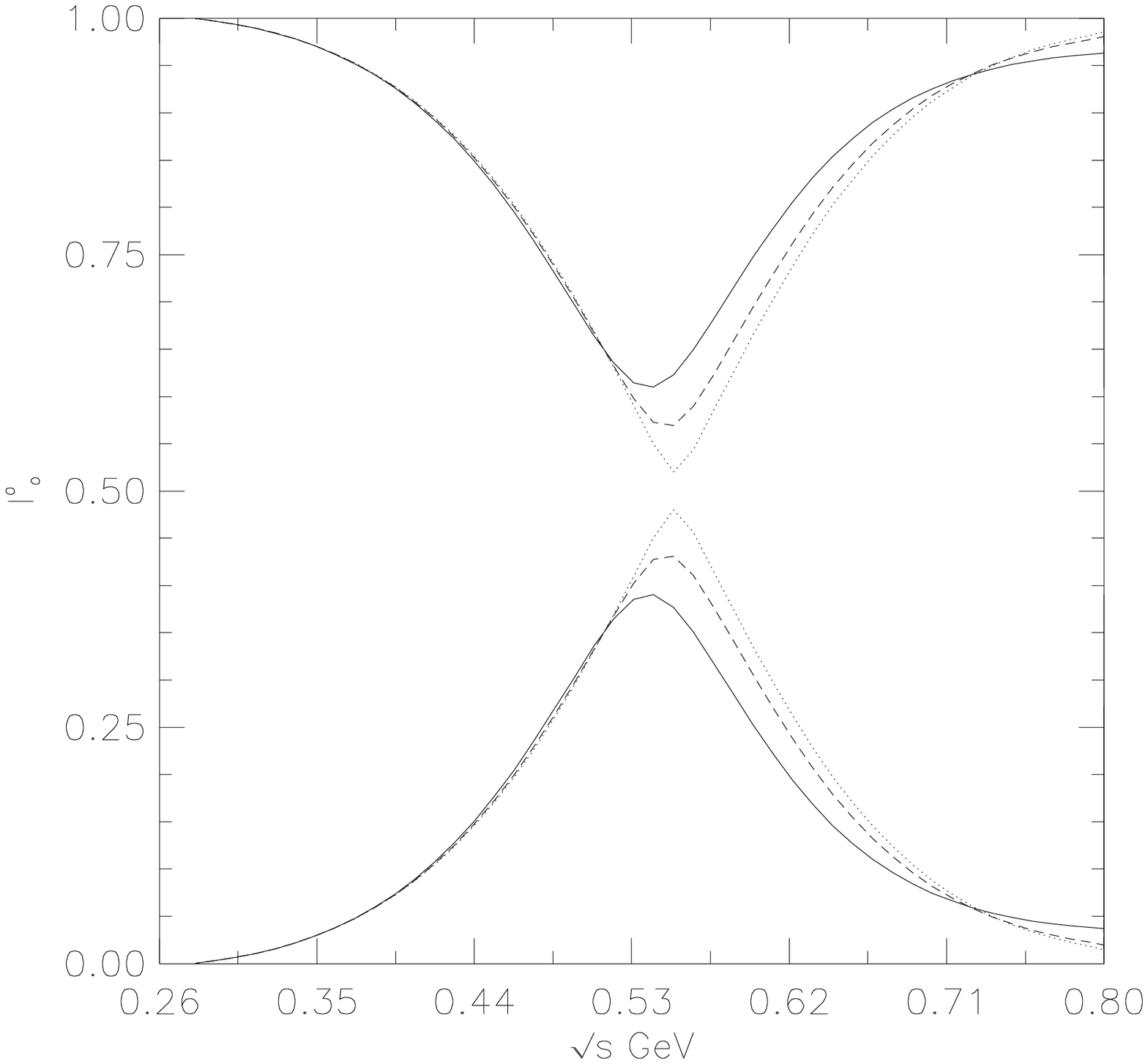}
\begin{itemize}
\item[Fig. 12]
{ Estimated imaginary part $I^0_0$.
Upper curves $+$ sign in front of the square root
in (\ref{eq:Ima}), lower curves, $-$ sign.
Solid line:
$G^\prime=380~MeV$. Dashed line: $G^\prime=440~MeV$. Dotted line:
$G^\prime=470~MeV$.}
\end{itemize}
\end{center}
\noindent

\begin{center}
\epsfxsize=400pt
\ \epsfbox{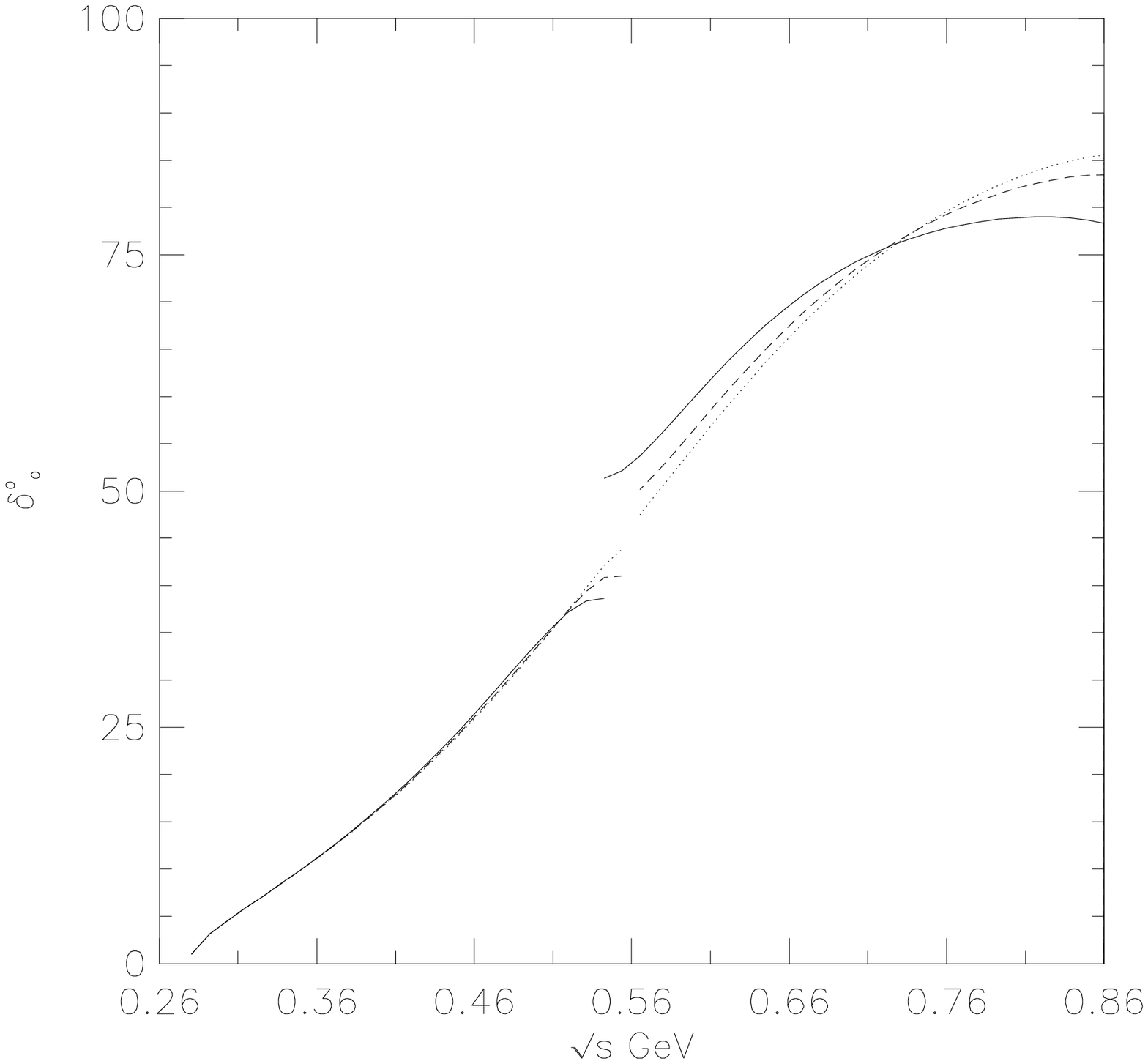}
\begin{itemize}
\item[Fig. 13]
{Estimated phase shift $\delta^0_0$. Solid line:
 $G^\prime=380~MeV$. Dashed line:
$G^\prime=440~MeV$. Dotted line:
$G^\prime=470~MeV$.}
\end{itemize}
\end{center}
\noindent
\end{document}